\documentclass[aps,prd,twocolumn,showpacs,preprintnumbers,nofootinbib,amsmath,amssymb]{revtex4}

\usepackage{amsmath}
\usepackage{color}
\usepackage{bm}
\usepackage{dcolumn}
\usepackage{graphicx}
\usepackage{multirow}

\newcommand{\beq}{\begin{eqnarray}}
\newcommand{\eeq}{\end{eqnarray}}

\newcommand{\rmd}{\ensuremath{\mathrm{d}}}

\allowdisplaybreaks  

\begin{document}

\title{$\theta$ dependence of 4D $SU(N)$ gauge theories in the
  large-$N$ limit}

\author{Claudio Bonati}
\email{claudio.bonati@df.unipi.it}

\author{Massimo D'Elia}
\email{massimo.delia@unipi.it}

\author{Paolo Rossi}
\email{paolo.rossi@unipi.it}

\author{Ettore Vicari}
\email{ettore.vicari@unipi.it}

\affiliation{Dipartimento di Fisica, Universit\`a di Pisa and INFN, Sezione di
Pisa, Largo Pontecorvo 3, 56127 Pisa, Italy}

\date{\today}

\begin{abstract}

We study the large-$N$ scaling behavior of the $\theta$ dependence of
the ground-state energy density $E(\theta)$ of four-dimensional (4D)
$SU(N)$ gauge theories and two-dimensional (2D) $CP^{N-1}$ models,
where $\theta$ is the parameter associated with the Lagrangian
topological term.  We consider its $\theta$ expansion around
$\theta=0$, $E(\theta)-E(0) = {1\over 2}\chi \,\theta^2 ( 1 + b_2
\theta^2 + b_4\theta^4 +\cdots)$ where $\chi$ is the topological
susceptibility and $b_{2n}$ are dimensionless coefficients.  We focus
on the first few coefficients $b_{2n}$, which parametrize the
deviation from a simple Gaussian distribution of the topological
charge at $\theta=0$.

We present a numerical analysis of Monte Carlo simulations of 4D
$SU(N)$ lattice gauge theories for $N=3,\,4,\,6$ in the presence of an
imaginary $\theta$ term. The results provide a robust evidence of the
large-$N$ behavior predicted by standard large-$N$ scaling arguments,
i.e. $b_{2n}= O(N^{-2n})$.  In particular, we obtain
$b_2=\bar{b}_2/N^2 + O(1/N^4)$ with $\bar{b}_2=-0.23(3)$.  We also
show that the large-$N$ scaling scenario applies to 2D $CP^{N-1}$
models as well, by an analytical computation of the leading large-$N$
$\theta$ dependence around $\theta=0$.

\end{abstract}

\maketitle

\section{Introduction}\label{sec:intro}

Some of the most intriguing properties of 4D $SU(N)$ gauge theories
are those related to the dependence on the $\theta$ parameter
associated with a topological term in the (Euclidean) Lagrangian
\begin{equation}\label{lagrangian}
\mathcal{L}_\theta  = \frac{1}{4} F_{\mu\nu}^a(x)F_{\mu\nu}^a(x)
- i \theta q(x)\ ,
\end{equation}
where $q(x)$ is the topological charge density,
\begin{equation}\label{topchden}
q(x)=\frac{g^2}{64\pi^2} 
\epsilon_{\mu\nu\rho\sigma} F_{\mu\nu}^a(x) F_{\rho\sigma}^a(x)\ .
\end{equation}
The dependence on $\theta$ vanishes perturbatively, therefore it is
intrinsically nonperturbative~\cite{book1,book2,book3}.

The recent renewed activity in the study of the topological properties
of gauge field theories, and of $\theta$ dependence in particular, has
been triggered by two different motivations. From the purely
theoretical point of view $\theta$-related topics naturally appear in
such disparate conceptual frameworks as the semiclassical methods
\cite{Gross:1980br, Schafer:1996wv, Thomas:2011ee, Unsal:2012zj,
  Poppitz:2012nz, Anber:2013sga, Kharzeev:2015xsa}, the expansion in
the number of colors \cite{Witten:1980sp, DiVecchia:1980ve,
  Luscher:1978rn, D'Adda:1978uc, Witten:1978bc,AA-80,Campostrini:1991kv,
  CR-92}, the holographic approach \cite{Witten:1998uka,
  Parnachev:2008fy, Bigazzi:2014qsa, Bigazzi:2015bna} and the lattice
discretization (see, e.g., \cite{Vicari:2008jw} for a review of the
main results).  From the phenomenological point of view, the
nontrivial $\theta$ dependence is related to the breaking of the axial
$U_A(1)$ symmetry and related issues of the hadronic
phenomenology~\cite{Witten:1979vv, Veneziano:1979ec, Shore:2007yn},
such as the $\eta'$ mass. Moreover, it is related to the axion physics
(see, e.g., \cite{Kim:2008hd} for a recent review), put forward to
provide a solution of the strong CP problem \cite{Peccei:1977hh,
  Peccei:1977ur, Wilczek:1977pj, Weinberg:1977ma}, i.e. to explain the
fact that the experimental value of $\theta$ is compatible with zero,
with a very small bound $|\theta| \lesssim 10^{-10}$ from neutron
electric dipole measurements \cite{Baker:2006ts}. Axions are also
natural dark matter candidates \cite{Preskill:1982cy, Abbott:1982af,
  Dine:1982ah} and, given the absence of SUSY signals from accelerator
experiments, this is becoming one of the most theoretically appealing
possibility.

The ground-state energy density of 4D $SU(N)$ gauge theories is an
even function of $\theta$. It is expected to be analytic at
$\theta=0$, thus it can be expanded in the form
\begin{equation}\label{thdep}
E(\theta)-E(0)=\frac{1}{2}\chi\theta^2\big(1+b_2\theta^2+b_4\theta^4+\cdots\big)\ ,
\end{equation}
where $\chi$ is the topological susceptibility, and the dimensionless
coefficients $b_{2n}$ parametrize the non-quadratic part of the
$\theta$ dependence.  They are related to the cumulants of the
topological charge distribution at $\theta=0$; in particular $b_{2n}$
quantify the deviations from a simple Gaussian distribution.  Standard
large-$N$ arguments predict the large-$N$
behavior~\cite{Witten:1980sp, Witten:1998uka,Vicari:2008jw}
\begin{eqnarray}\label{largeNsun}
&&\chi(N)=\bar{\chi}+O(N^{-2}), \label{largenchi}\\
&&b_{2n}(N)=\bar{b}_{2n}N^{-2n}+O(N^{-2n-2}).
\label{largenbn}
\end{eqnarray}
Since $\chi$ and $b_{2n}$ can not be computed
analytically, these large-$N$ scaling relations can be only tested
numerically.

Earlier studies have mainly focused on the investigation of the
large-$N$ scaling of the topological susceptibility, reporting a good
agreement with the corresponding large-$N$ expectations (see, e.g.,
\cite{Lucini:2001ej, DelDebbio:2002xa,LTW-05}). Instead, the numerical
determination of the higher-order coefficients of the $\theta$
expansion turns out to be a difficult numerical challenge. Most
efforts have been dedicated to the $SU(3)$
case~\cite{DelDebbio:2002xa,D'Elia:2003gr,Giusti:2007tu,Panagopoulos:2011rb,
  Ce:2015qha,Bonati:2015sqt}, reaching a precision corresponding to a
relative error below 10\% only recently~\cite{Bonati:2015sqt}.  Some
higher-$N$ results were reported in Ref.~\cite{DelDebbio:2002xa},
presenting a first attempt to investigate the large-$N$ scaling of
$b_2$; the numerical precision that could be reached was however quite
limited, with a signal for $SU(4)$ at two standard deviation from zero
and only an upper bound for $SU(6)$.

In order to further support the evidence of the large-$N$ scaling
scenario beyond the quadratic term of the expansion of the
ground-state energy dentity (\ref{thdep}), we investigate the scaling
of the higher-order terms, in particular those associated with $b_2$
and $b_4$.  From the computational point of view, the most convenient
method to perform such an investigation exploits Monte Carlo
simulations of $SU(N)$ gauge theories in the presence of an imaginary
$\theta$ angle, which are not plagued by the sign problem.  Their
analysis allows us to obtain accurate estimates of the coefficients of
the expansion around $\theta=0$.  Analogous methods based on
computations at imaginary values of $\theta$ have been already
employed in some numerical studies of the $SU(3)$ gauge
theory~\cite{aoki_1,Panagopoulos:2011rb, D'Elia:2012vv,
  D'Elia:2013eua, Bonati:2015sqt} and $CP^{N-1}$ models
\cite{Azcoiti:2002vk, Alles:2007br, Alles:2014tta}.

2D $CP^{N-1}$ models share with 4D $SU(N)$ gauge theories many
physically interesting properties, like asymptotic freedom, dynamical
mass generation, confinement, instantons and $\theta$ dependence;
moreover their large-$N$ expansion can be studied by analytical
methods~\cite{Luscher:1978rn, D'Adda:1978uc, Witten:1978bc,AA-80,
  Campostrini:1991kv,CR-92}.  As a consequence they are an attractive
laboratory where to test theoretical ideas that might turn out to be
applicable to QCD. An expansion of the form Eq.~\eqref{thdep} applies
also to the $\theta$ dependence of 2D $CP^{N-1}$ models.  Similarly to
4D $SU(N)$ gauge theories, large-$N$ scaling arguments predict the
large-$N$ behavior $\chi\approx c\,N^{-1}$ and $b_{2n}\approx
\bar{b}_{2n}\,N^{-2n}$.  These large-$N$ scaling behaviors are
confirmed by explicit analytical computations,
see~\cite{Luscher:1978rn, D'Adda:1978uc, Witten:1978bc,
  Campostrini:1991kv,DelDebbio:2006yuf}.  In this paper, following the
approach introduced in \cite{Rossi:2016uce}, we present a systematic
and easily automated way of computing the leading large-$N$ terms of
$b_{2n}$.

The paper is organized as follows. In Section~\ref{sec:sun} we present
the results obtained for the case of the 4D $SU(N)$ gauge theories:
first we discuss the numerical setup used and the reasons for some
specific algorithmic choices adopted, then we present the physical
results obtained. In Section~\ref{sec:cpn} the case of the 2D
$CP^{N-1}$ models is discussed and a determination of the leading
order large-$N$ expansion for the coefficients $b_{2n}$ is
presented. Finally, in Section~\ref{sec:concl}, we draw our
conclusions. In appendices some technical details are examined
regarding a comparison between smoothing algorithms in $SU(6)$
(App.~\ref{sec:cgf}) and an attempt to reduce the autocorrelation time
using a parallel tempering algorithm (App.~\ref{sec:pt}).  Tables of
numerical data are reported in App.~\ref{sec:data}.

\section{Large $N$ in 4D $SU(N)$ gauge theories}\label{sec:sun}

\subsection{Numerical setup}\label{sec:numset}

The traditional procedure that has been used in past to compute the
coefficients entering Eq.~\eqref{thdep} consists in relating them to
the fluctuations of the topological charge $Q\equiv \int q(x) \rmd^d
x$ at $\theta=0$. The first few coefficients of the expansion can
indeed be written as (see e.g. \cite{Vicari:2008jw})
\begin{align}
& \chi = \frac{\langle Q^2 \rangle_{\theta=0}}{\mathcal{V}}, \label{chi} \\
& b_2=-\frac{\langle Q^4\rangle_{\theta=0}-
         3\langle Q^2\rangle^2_{\theta=0}}{12\langle Q^2\rangle_{\theta=0}}, \label{b2}\\
& b_4=\frac{\left[ \langle Q^6\rangle-15\langle Q^2\rangle \langle Q^4\rangle 
+30\langle Q^2\rangle ^3\right]_{\theta=0}}{360 \langle Q^2\rangle_{\theta=0}} ,\label{b4}
\end{align}
etc., where $\mathcal{V}$ is the 4D volume and all the averages are
computed using the action with $\theta=0$. While this method is
obviously correct from the theoretical point of view, it is
numerically inefficient for the determination of the $b_{2n}$
coefficients.  Indeed fluctuation observables are not self-averaging
\cite{MBH} and, in order to keep a constant signal to noise ratio, one
has to dramatically increase the statistics of the simulations when
increasing the volume (see e.g. \cite{Bonati:2015sqt} for a numerical
example).  As a consequence it is extremely difficult to keep finite
size effects under control and to extract the infinite-volume limit.

To avoid this problem, one can introduce a source term in the action,
which allows us to better investigate the response of the system. This
can be achieved by performing numerical simulations at imaginary
values of the $\theta$ angle, $\theta\equiv -i\theta_I$, in order to
maintain the positivity of the path integral measure and avoid a sign
problem, and study for example the behaviour of $\langle
Q\rangle_{\theta_I}$ as a function of $\theta_I$. It is indeed easy to
verify that \cite{Panagopoulos:2011rb}
\begin{equation}\label{Qcum1th}
\frac{\langle Q\rangle_{\theta_I}}{\mathcal{V}}=\chi\theta_I(1-2b_2\theta_I^2+
3b_4\theta_I^4+\ldots )\ .
\end{equation}
Also higher cumulants of the topological charge distribution (for
which relations analogous to Eq.~\eqref{Qcum1th} exist) can be used
for this purpose, however the numerical precision quickly degrades for
higher cumulants. Nevertheless, since the computation of these higher
cumulants does not require any additional CPU time, the optimal
strategy seems to be to perform a common fit to a few of the lowest
cumulants of the topological charge \cite{Bonati:2015sqt} (of course
by taking into account the correlation between them).

After this general introduction to motivate the computational strategy
adopted, we describe the details of the discretization setup.  For the
$SU(3)$ case we use results already reported in the literature, while
new simulations are performed for the $SU(4)$ and $SU(6)$ cases. The
lattice action used in the sampling of the gauge configurations is
\begin{equation}\label{latticeaction}
S[U] = S_W[U] - \theta_{L} Q_L[U] \, ,
\end{equation}
where $S_W[U]$ is the standard Wilson plaquette action
\cite{Wilson:1974sk} and $Q_L = \sum_x q_L(x)$.  For the topological
charge density we adopt the discretization \cite{DiVecchia:1981qi,
  DiVecchia:1981hh}:
\begin{equation}\label{qlattice}
q_L(x) = -\frac{1}{2^9 \pi^2} 
\sum_{\mu\nu\rho\sigma = \pm 1}^{\pm 4} 
{\tilde{\epsilon}}_{\mu\nu\rho\sigma} \hbox{Tr} \left( 
\Pi_{\mu\nu}(x) \Pi_{\rho\sigma}(x) \right) \; ,
\end{equation}
where $\Pi_{\mu\nu}$ is the plaquette,
$\tilde{\epsilon}_{\mu\nu\rho\sigma}$ coincides with the usual
Levi-Civita tensor for positive entries and it is extended to negative
ones by ${\tilde{\epsilon}}_{\mu\nu\rho\sigma} =
-{\tilde{\epsilon}}_{(-\mu)\nu\rho\sigma}$ and complete antisymmetry.
The discretization Eq.~\eqref{qlattice} of the topological charge
density makes the total action in Eq.~\eqref{latticeaction} linear in
each gauge link, thus enabling the adoption of standard efficient
update algorithms, like heat-bath and overrelaxation, a fact of
paramount importance, since we have to deal with the strong critical
slowing down of the topological modes~\cite{DelDebbio:2002xa}.

A practical complication is due to the fact that the discretization of
the topological charge density induces a finite renormalization of
$q(x)$~\cite{Campostrini:1988cy} and thus of $\theta$.  Denoting this
renormalization constant by $Z$, we thus have
\begin{equation}
\theta_I=Z\theta_L,
\label{thetairen}
\end{equation}
where $\theta_L$ is the numerical value that is used in the actual
simulation. Two different strategies can be used to cope with this
complication: in one case $Z$ is computed separately and
Eq.~\eqref{Qcum1th} can then be directly used (see
\cite{Panagopoulos:2011rb} for more details).  Another possibility
consists in rewriting Eq.~\eqref{Qcum1th}, and the analogous equations
for the higher cumulants, directly in term of $\theta_L$, in such a
way that by performing a common fit to the cumulants it is possible to
evaluate both $Z$ and the parameters appearing in Eq.~\eqref{thdep}
(see \cite{Bonati:2015sqt} for more details). In our numerical work we
adopt the second strategy, that turn out to be slightly more efficient
from the numerical point of view. All results that we present are
obtained by analyzing the $\theta_I$ dependence of the first four
cumulants of the topological charge distribution.

In order to avoid the appearance of further renormalization factors,
the topological charge is measured on smoothed configurations. The
smoothing procedure adopted uses the standard cooling
technique~\cite{Berg:1981nw, Iwasaki:1983bv, Itoh:1984pr,
  Teper:1985rb, Ilgenfritz:1985dz}, which is the computationally
cheapest procedure (especially for large values of $N$).  Cooling is
implemented \emph{\`a la} Cabbibbo-Marinari, using the $N(N-1)/2$
diagonal $SU(2)$ subgroups of $SU(N)$, and we follow
\cite{DelDebbio:2002xa} in defining the measured topological charge
$Q$ by
\begin{equation}\label{Q_round}
Q=\mathrm{round}\left(\alpha\, Q_L^{\rm smooth} \right)\ , 
\end{equation} 
where $\mathrm{round}(x)$ is the integer closest to $x$ and the coefficient 
$\alpha$ is the value that minimize
\begin{equation}
\left\langle \left( \alpha\, Q_L^{\rm smooth} - \mathrm{round}
\left[\alpha\, Q_L^{\rm smooth} \right]\right)^2\right\rangle \ .
\end{equation}
This procedure is introduced in order to avoid the necessity for
prolongated cooling, and in fact we observe no significant differences
in the results obtained by using a number of cooling steps between $5$
and $25$, while more than $100$ cooling steps would be needed to reach
a plateau using just $Q_L$ instead of the $Q$ defined by
Eq.~\eqref{Q_round}. At finite lattice spacing, the two definitions
(rounded vs. non-rounded) can lead to different results corresponding
to different lattice artefacts, however it has been shown that the
same continuum limit is reached in the two
cases~\cite{Bonati:2015sqt}.  The results that we present in the
following are obtained using $15$ cooling steps and the definition of
$Q$ in Eq.~\eqref{Q_round}.  We also mention that the results of this
cooling procedure are compatible with those of other approaches
proposed in the literature, see \cite{Vicari:2008jw, Bonati:2014tqa,
  Cichy:2014qta, Namekawa:2015wua, Alexandrou:2015yba} and
App.~\ref{sec:cgf}.

Seven $\theta_L$ values are typically used in the simulations, going
from $\theta_L=0$ to $\theta_L=12$ with steps $\Delta\theta_L=2$; when
expressed in term of the renormalized parameter $\theta_I=Z\theta_L$
this range of $\theta_L$ values corresponds (for the couplings used in
this work) to $\theta_I\lesssim 1.8$. We verify that this range of
values is large enough to give a clear signal, but not so large to
introduce systematic errors. The results of all tests performed using
a smaller interval of $\theta_L$ values give perfectly compatible
results.

For the update we use a combination of standard heat-bath
\cite{Creutz:1980zw, Kennedy:1985nu} and overrelaxation
\cite{Creutz:1987xi} algorithms, implemented \emph{\`a la}
Cabibbo-Marinari \cite{Cabibbo:1982zn} using all the $N(N-1)/2$
diagonal $SU(2)$ subgroups of $SU(N)$.  The topological charge is
evaluated every 10 update steps, one update step being composed of a
heath-bath and five overrelaxation updates for all the links of the
lattice, updated in a mixed checkerboard and lexicographic order. The
total statistic acquired for each coupling value is typically of
$O(10^6)$ measures.

\subsection{Numerical results}\label{sec:numres}

\begin{figure}
\includegraphics[width=0.92\columnwidth, clip]{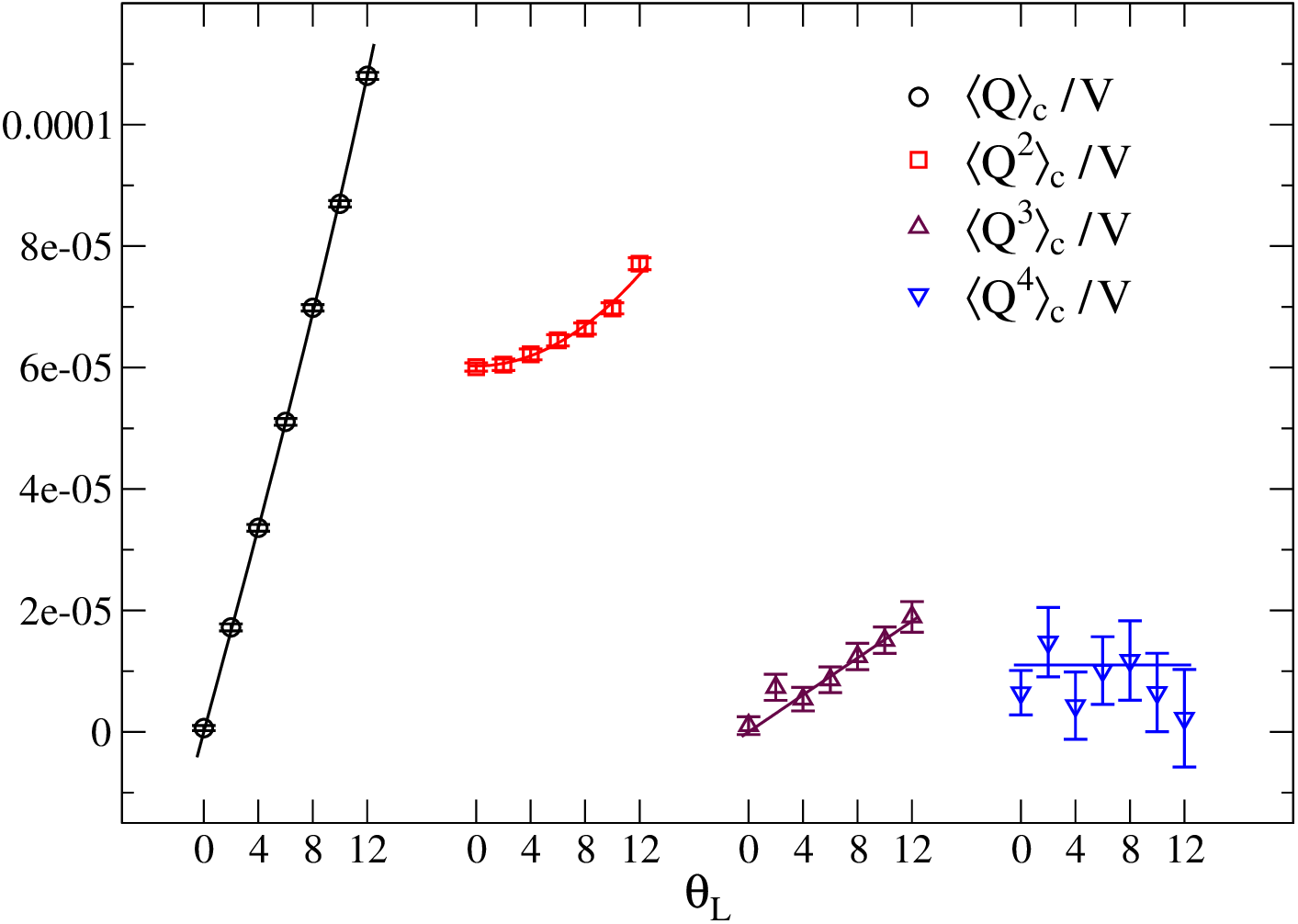}
\caption{An example of the global fit procedure, with a truncation including
$O(\theta_L^4)$ terms: data refer to the $14^4$ lattice at coupling
$\beta=11.008$ for the $SU(4)$ gauge theory.  Continuous lines are the result
of a combined fit of the first four cumulants.}
\label{fig:glob}
\end{figure}

In order to apply the analytic continuation method in an actual
computation, it is necessary to truncate the expansion in
Eq.~\eqref{Qcum1th} (or, which is the same, in Eq.~\eqref{thdep}) in
order to fit the numerical data.  We actually perform a global fit to
the first four cumulants which, when rewritten in terms of $\theta_L$,
read
\begin{equation}\label{eq:cum_theta_L}
\begin{split}
\frac{\langle Q \rangle}{\mathcal{V}}&=
\chi Z \theta_L (1 - 2 b_2 Z^2 \theta_L^2 + 3 b_4 Z^4 \theta_L^4 + \dots)\, , \\
\frac{\langle Q^2 \rangle_c}{\mathcal{V}}&=  
\chi (1 - 6 b_2 Z^2 \theta_L^2 + 15 b_4 Z^4 \theta_L^4 + \dots)\, , \\
\frac{\langle Q^3 \rangle_c}{\mathcal{V}}&=  
\chi (- 12 b_2 Z \theta_L + 60 b_4 Z^3 \theta_L^3 + \dots)\, , \\ 
\frac{\langle Q^4 \rangle_c}{\mathcal{V}}&=
\chi (- 12 b_2 + 180 b_4 Z^2 \theta_L^2 + \dots)\, .
\end{split}
\end{equation}
An example of such global fit, with a truncation including up to
$O(\theta_L^4)$ terms in the ground state energy density (i.e. setting
$b_4 = 0$), is reported in Fig.~\ref{fig:glob} for the case of the
$SU(4)$ gauge theory.
  
\begin{figure}
\includegraphics[width=0.92\columnwidth, clip]{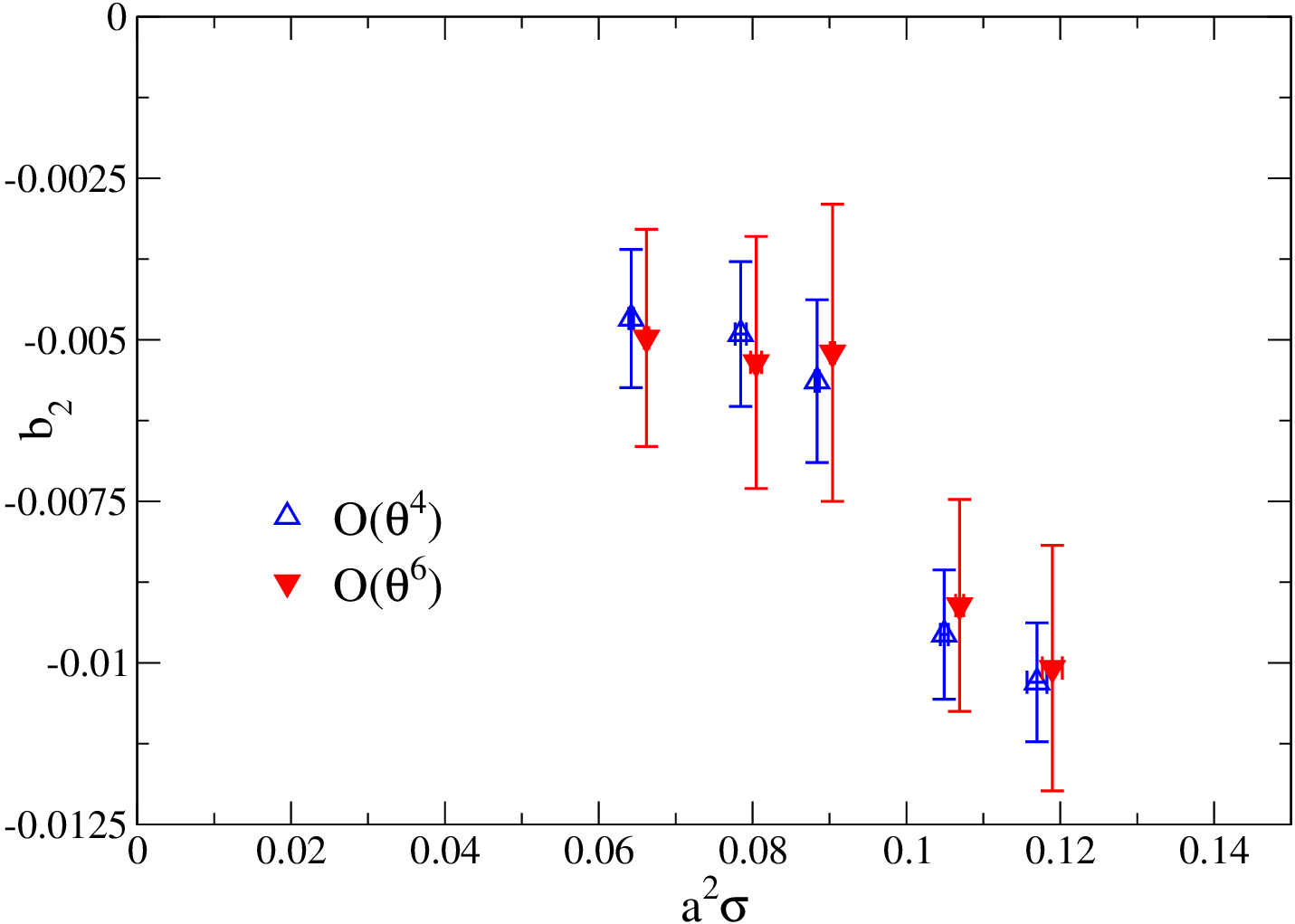}
\caption{Comparison of the results obtained for $b_2$ in $SU(6)$ using
different truncations of Eq.~\eqref{thdep}.}
\label{fig:sys}
\end{figure}

To quantify the systematic error associated with this procedure we
consider two different truncations: in one case all the terms of
Eq.~\eqref{thdep} up to $O(\theta^6)$ are retained (i.e. up to $b_4$),
while in the other case a truncation up to $O(\theta^4)$ (i.e. up to
$b_2$) is used. Both truncations nicely fit the numerical data and the
estimates of the coefficient $b_4$ turn out to be compatible with zero
in all the cases. This is not surprising, since even for $SU(3)$ only
upper bounds on $|b_4|$ exist (see, e.g., \cite{Panagopoulos:2011rb,
  Bonati:2015sqt}) and its value is expected to approach zero very
quickly as the number of colors is increased, see
Eq.~\eqref{largeNsun}. We verify that the values of $Z, \chi$ and
$b_2$ obtained by using the two different truncations are perfectly
compatible with each other, indicating that no sizable systematic
error is introduced by the truncation procedure, see the example in
Fig.~\ref{fig:sys}.  For this reason we decide to use the
$O(\theta^4)$ truncation to estimate $Z$, $\chi$ and $b_2$, while the
$O(\theta^6)$ truncation is obviously needed to obtain an upper bound
for $|b_4|$.  Possible further systematic errors are checked by
varying the fitted range of $\theta_L$ and verifying the stability of
the fit parameters.

\begin{figure}
\includegraphics[width=0.92\columnwidth, clip]{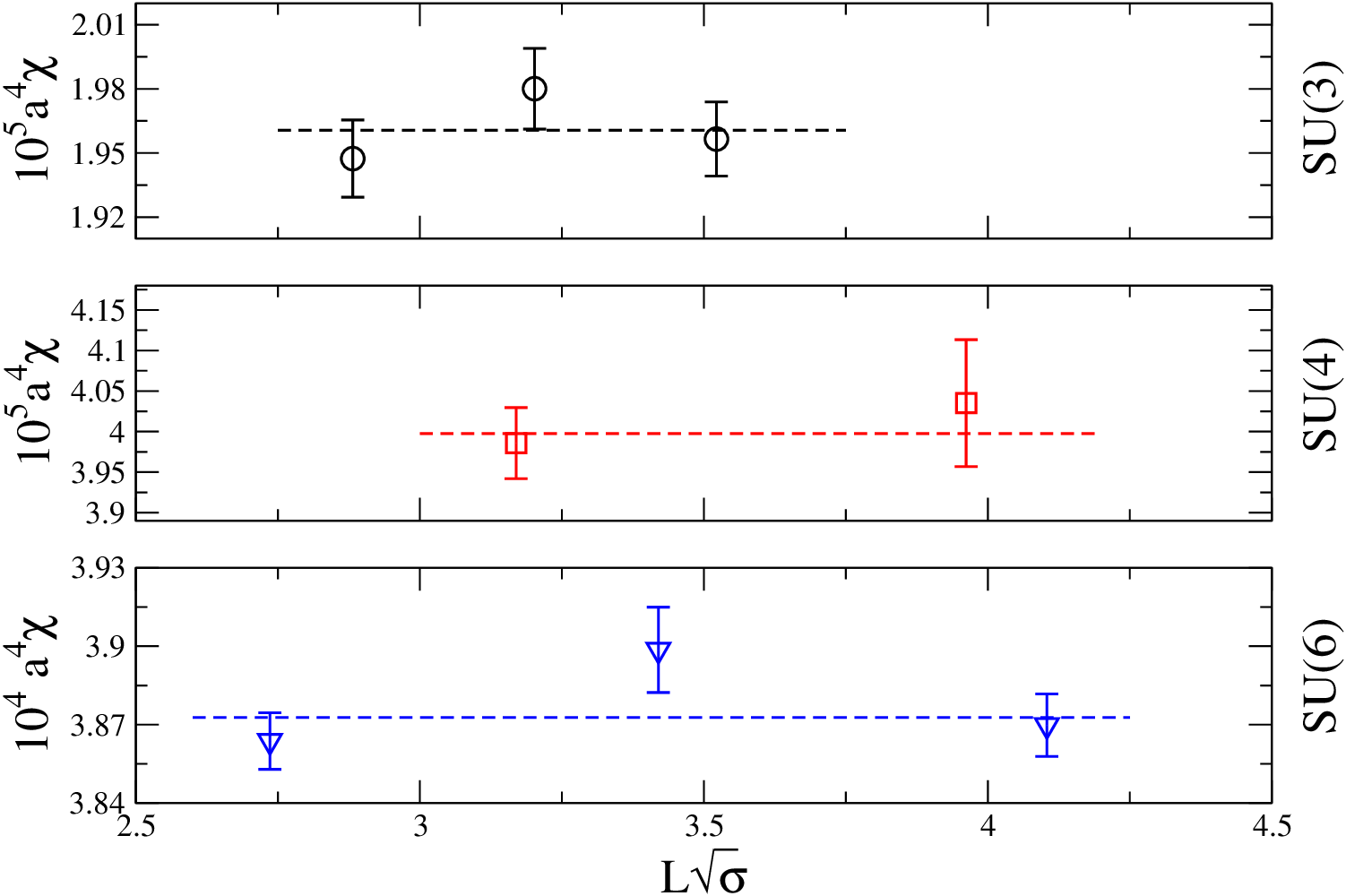}
\caption{Dependence of $\chi$ (in lattice units) on the lattice size. From top
to bottom results are displayed for: $SU(3)$ at coupling $\beta=6.2$ (from
\cite{Bonati:2015sqt}), $SU(4)$ at $\beta=11.104$ and $SU(6)$ at
$\beta=24.500$. Horizontal dashed lines are the results of fits to constant and
are plotted in order to better appreciate the absence of statistically
significant deviations.}
\label{fig:chi_fs}
\end{figure}

\begin{figure}
\includegraphics[width=0.92\columnwidth, clip]{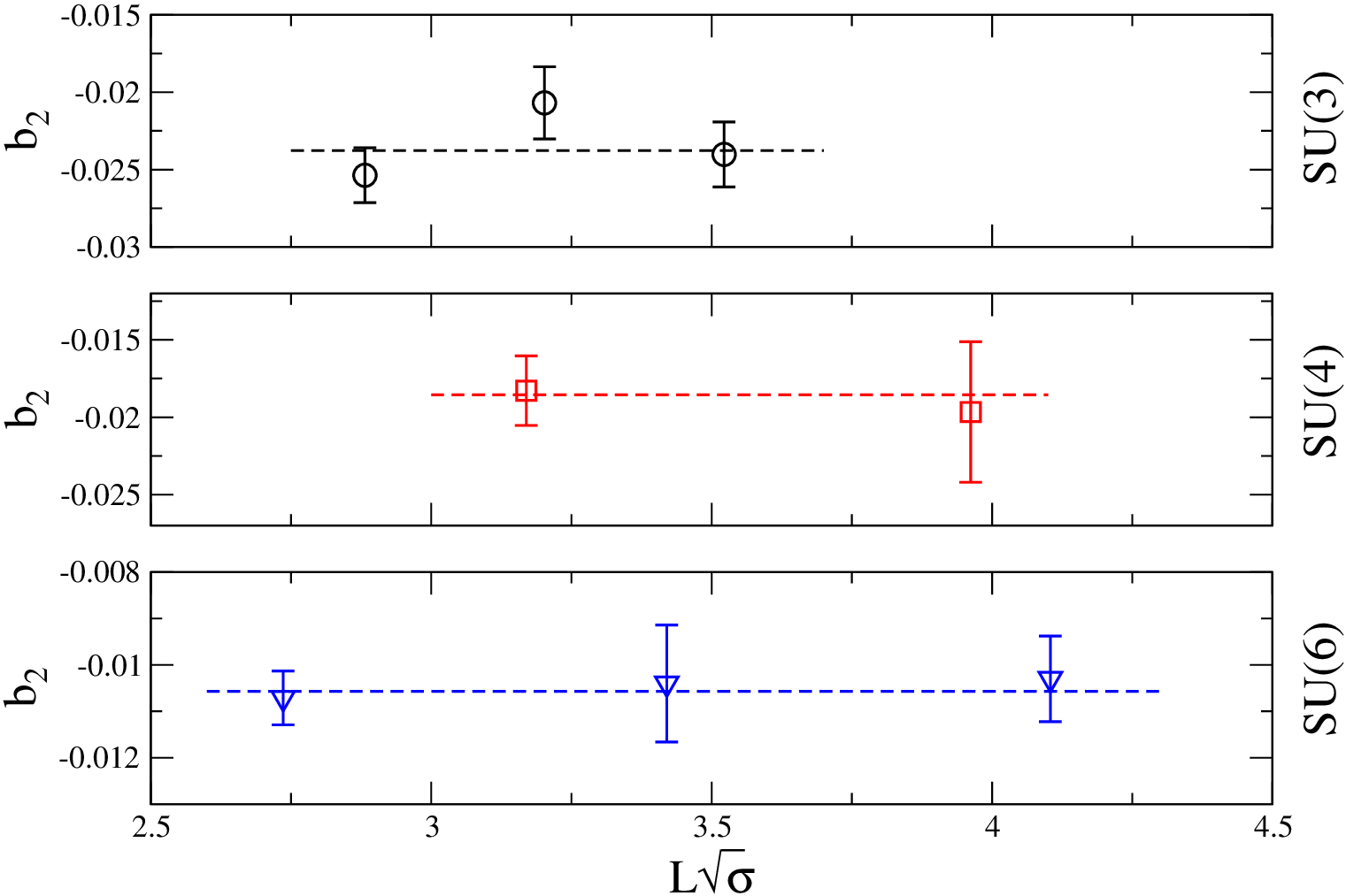}
\caption{Dependence of $b_2$ on the lattice size. From top to bottom results
are displayed for: $SU(3)$ at coupling $\beta=6.2$ (from
\cite{Bonati:2015sqt}), $SU(4)$ at $\beta=11.104$ and $SU(6)$ at
$\beta=24.500$. Horizontal dashed lines are the results of fits to constant and
are plotted in order to better appreciate the absence of statistically
significant deviations.}
\label{fig:b2_fs}
\end{figure}

\begin{figure}
\includegraphics[width=0.92\columnwidth, clip]{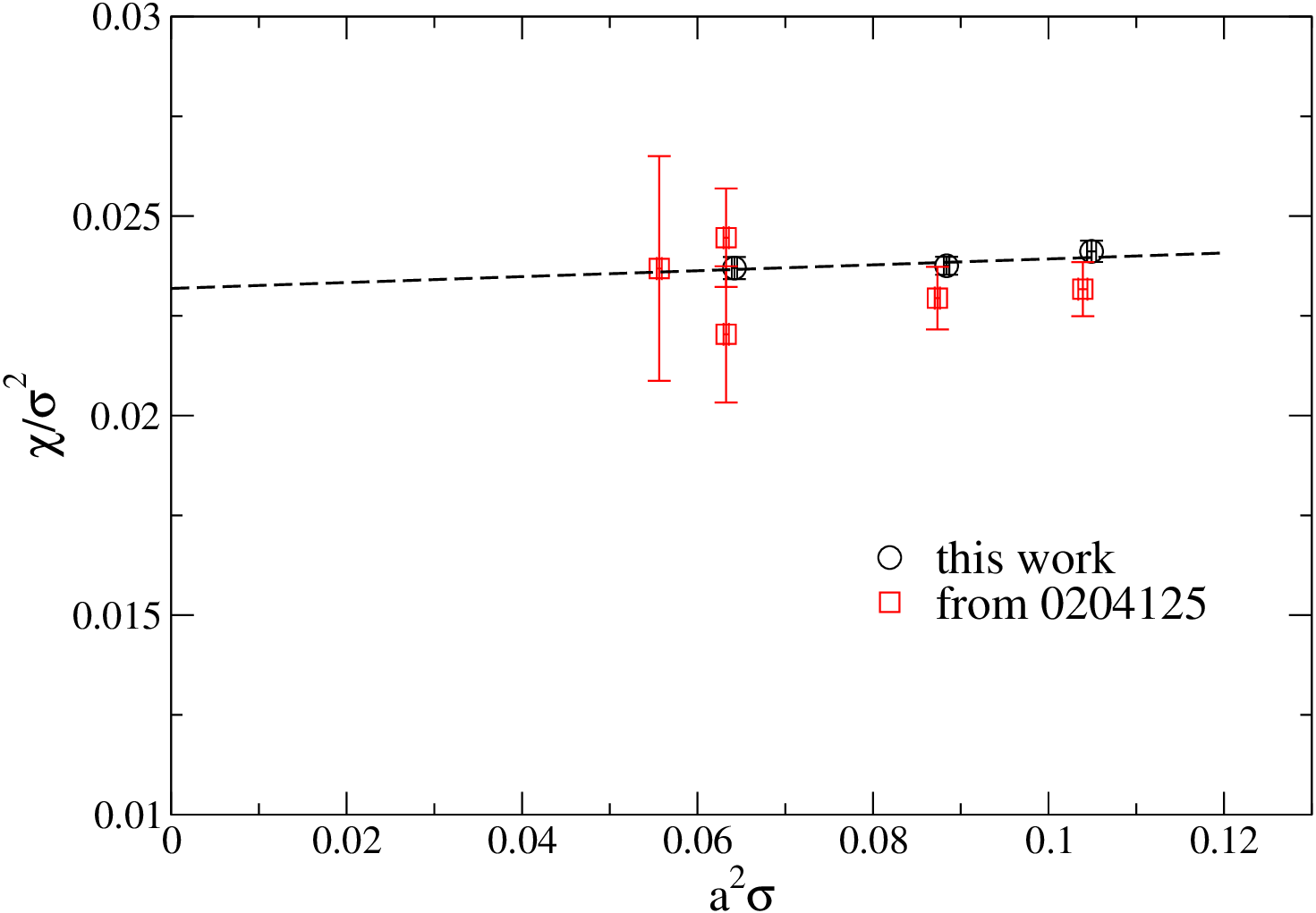}
\caption{Continuum limit of the dimensionless ratio $\chi/\sigma^2$ for $SU(6)$
gauge theory. The results obtained in this work are compared with the
determination of \cite{DelDebbio:2002xa} (data have been slightly shifted
horizontally to improve the readability).}
\label{fig:chisu6}
\end{figure}

\begin{table}
\begin{tabular}{l|l|l|l}
\hline \rule{0mm}{3mm}$N$ & $\quad \chi/\sigma^2$ & $\qquad b_2$ & $\qquad b_4$ \\
\hline 3   & 0.0289(13)      & $-$0.0216(15) & $\phantom{-}$0.0001(3)\\ 
\hline 4   & 0.0248(8)       & $-$0.0155(20) & $-$0.0003(3) \\
\hline 6   & 0.0230(8)       & $-$0.0045(15) & $-$0.0001(7) \\
\hline
\end{tabular}
\caption{Continuum extrapolated values for three, four and six colors. 
The value of $\chi/\sigma^2$ in $SU(3)$ was computed using data from 
\cite{Panagopoulos:2011rb}, while for $b_{2n}$ we used the value
reported in \cite{Bonati:2015sqt}.}
\label{tab:contvalues}
\end{table}

\begin{figure}
\includegraphics[width=0.92\columnwidth, clip]{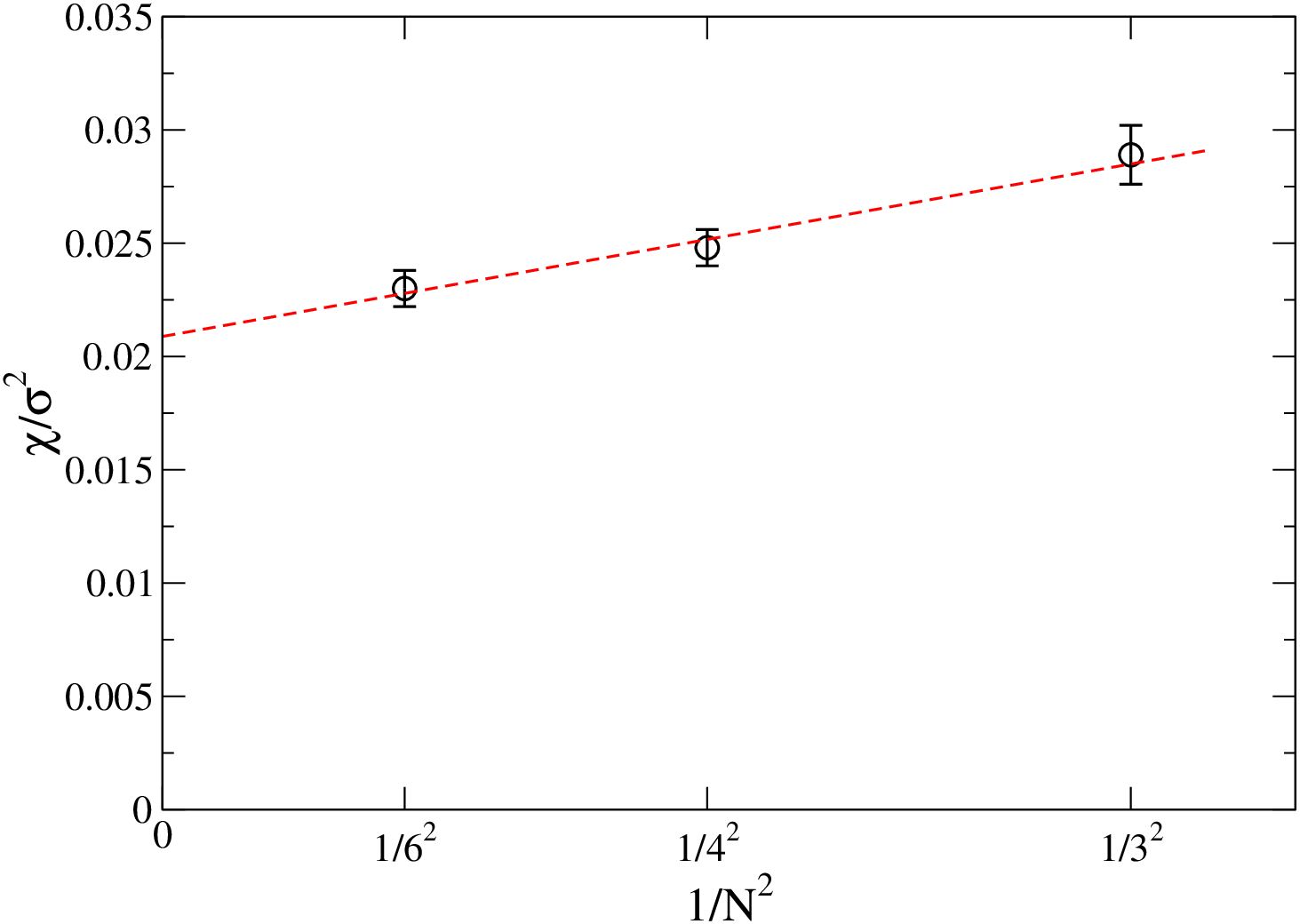}
\caption{Scaling of the dimensionless ratio $\chi/\sigma^2$ with the number of
colors. The dashed line is the result of a best fit with a linear functional
dependence.}
\label{fig:chilargeN}
\end{figure}

Hypercubic lattices of size $L\sqrt{\sigma}\gtrsim 3$ are used in all
cases: they are expected to be large enough to provide the
infinite-volume limit within the typical errors of our simulations
(see e.g.  \cite{DelDebbio:2002xa}). This is explicitly verified in
some test cases: for example the $SU(6)$ simulations at coupling
$\beta=24.500$ were replicated on lattices of size $L/a=8, 10, 12$ and
for the coupling $\beta=24.845$ on lattices with $L/a=10, 12, 16$; in
all cases no statistically significant volume dependence is observed,
see Figs.~\ref{fig:chi_fs}-\ref{fig:b2_fs}. The possibility of using
such large lattices in the determination of $b_2$ and higher cumulants
is a consequence of the numerical setup adopted, with simulations
performed at imaginary $\theta$ values.

Before starting to discuss our main subject, namely the determination
of $b_2$ and its large $N$ behavior, we show that our data reproduce
the well known large $N$ scaling of $\chi/\sigma^2$.  For $SU(3)$ we
use results already available in the literature (those reported in
Tab.~1 of \cite{Panagopoulos:2011rb}) and for the scale setting in the
$SU(4)$ and $SU(6)$ cases we used the determination of the string
tension reported in \cite{DelDebbio:2001sj}.  For $SU(4)$ we observe
no improvement with respect to the old results of
\cite{DelDebbio:2002xa}, since the final error on $\chi/\sigma^2$ is
dominated by the error on the string tension. This is also the case
for the final continuum result in the $SU(6)$ case, indeed we obtained
$\chi/\sigma^2|_{SU(6)}=0.0230(8)$ to be compared with the value
$0.0236(10)$ reported in Ref.~\cite{DelDebbio:2002xa}, however the
continuum extrapolation of the new results is much more solid, as
shown in Fig.~\ref{fig:chisu6}.

The continuum values of $\chi/\sigma^2$ for $N=3,4$ and $6$ are
reported in Tab.~\ref{tab:contvalues}, their scaling with $N$ is shown
in Fig.~\ref{fig:chilargeN} and the result of a linear fit in $1/N^2$
gives
\begin{equation}
\chi/\sigma^2|_{SU(\infty)}=0.0209(11)\ , 
\end{equation}
which slightly improves the previous result of
Ref.~\cite{Lucini:2001ej,DelDebbio:2002xa,LTW-05,Vicari:2008jw}.
Assuming the standard value $\sqrt{\sigma}=440$ Mev, we obtain
$\chi_{SU(\infty)}^{1/4} = 167(2)$ Mev.  As noted before, the dominant
source of error in $\chi/\sigma^2$ is the error on the string
tension. As a consequence, to improve this result it would be enough
to improve the precision of the $\sigma$ determination or to use
different observables to set the scale.  Since our main interest in
this work is the analysis of the higher order cumulants $b_{2n}$,
which are dimensionless, we have not pursued this investigation any
further.

\begin{figure}
\includegraphics[width=0.92\columnwidth, clip]{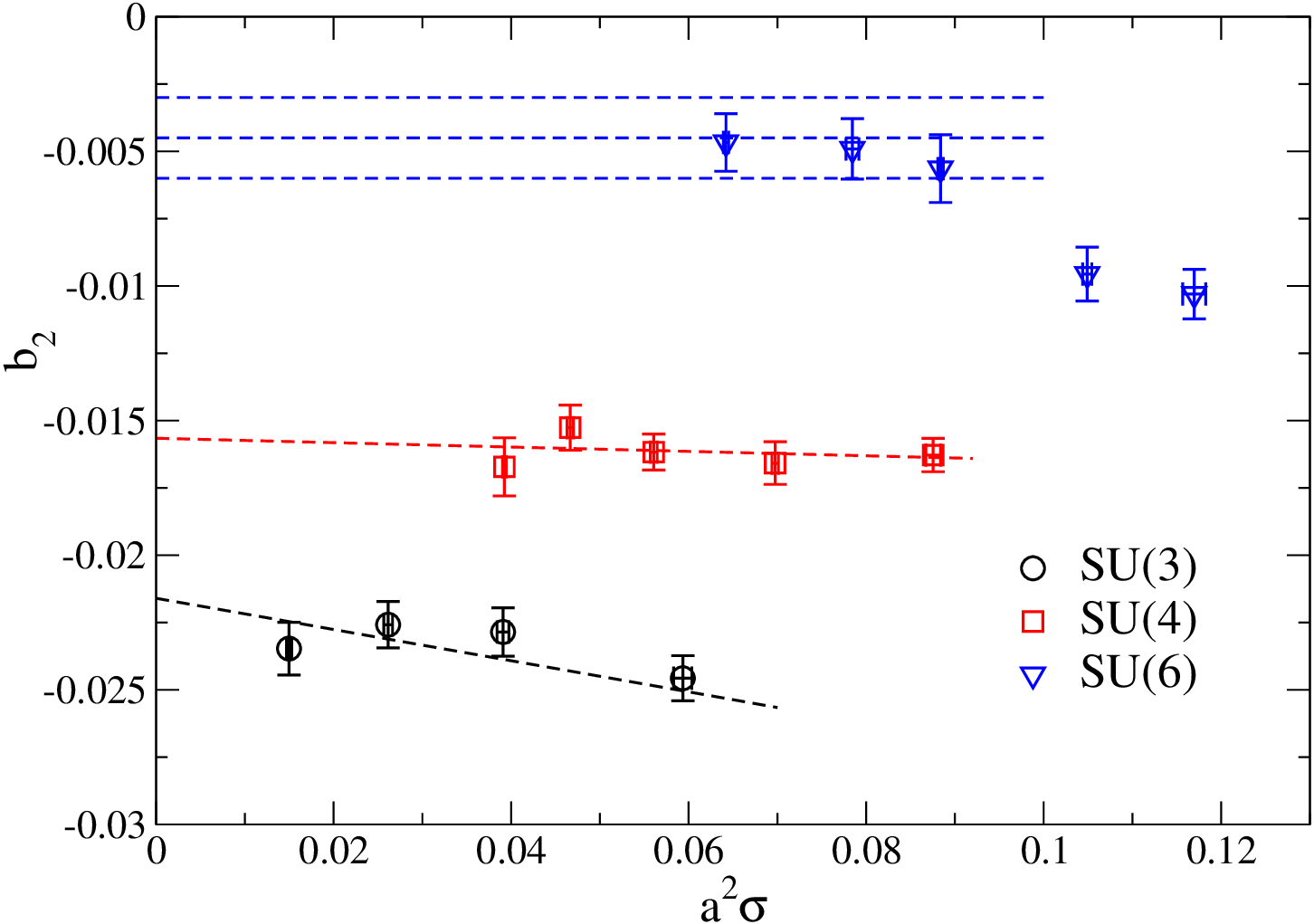}
\caption{Dependence of the $b_2$ values on the lattice spacing for the case of
three, four and six colors. See the main text for the details of the fitting
procedure.}
\label{fig:b2allN}
\end{figure}

In Fig.~\ref{fig:b2allN} the results obtained for $b_2$ with $N=3,4,6$ are
shown as a function of the (square of the) lattice spacing.  The values of
$a^2\sigma$ for the $SU(3)$ data have been computed using
$r_0\sqrt{\sigma}=1.193(10)$ from \cite{Niedermayer:2000yx} to plot the $b_2$
data from \cite{Bonati:2015sqt} (where $r_0$ was used to set the scale)
together with the new $SU(4)$ and $SU(6)$ data. For $SU(4)$, data are precise
enough to perform a linear fit in $a^2\sigma$ and check for the systematics of
the continuum extrapolation by varying the fit range; in particular the
final error reported in Tab.~\ref{tab:contvalues} takes into account also fits
obtained by excluding the data corresponding to the coarsest and to the finest
lattice spacings. (for $SU(3)$ we use the value obtained in
\cite{Bonati:2015sqt}, where a similar analysis was performed). For the case of
$SU(6)$ we could not reach lattice spacings as small as the ones used for
$SU(3)$ and $SU(4)$ due to the dramatic increase of the autocorrelation times
of the topological charge.  To boost the sampling we tried using parallel
tempering switches between different $\theta_L$ simulations but this did not
result in a significant improvement (see App.~\ref{sec:pt} for more details).
As a consequence, the analysis of the $SU(6)$ results can not be as
statistically accurate as those for $SU(3)$ and $SU(4)$. In spite of this, a
clear trend can be seen in the $SU(6)$ data shown in Fig.~\ref{fig:b2allN}:
$b_2$ flattens for $a^2\sigma\lesssim 0.1$, which is the region in which also
$SU(4)$ data show no significant dependence on the lattice spacing, and we use
the conservative estimate $b_2|_{SU(6)}=-0.0045(15)$, which is displayed in
Fig.~\ref{fig:b2allN} by the horizontal blue dashed lines.  For both $SU(4)$
and $SU(6)$ we increased significantly the precision of the $b_2$ determination
with respect to results available in the literature: the previous estimates
were indeed $b_2|_{SU(4)}=-0.013(7)$ and $b_2|_{SU(6)}=-0.01(2)$ from
\cite{DelDebbio:2002xa}, to be compared with the numbers reported in
Tab.~\ref{tab:contvalues}.
 
\begin{figure}
\includegraphics[width=0.92\columnwidth, clip]{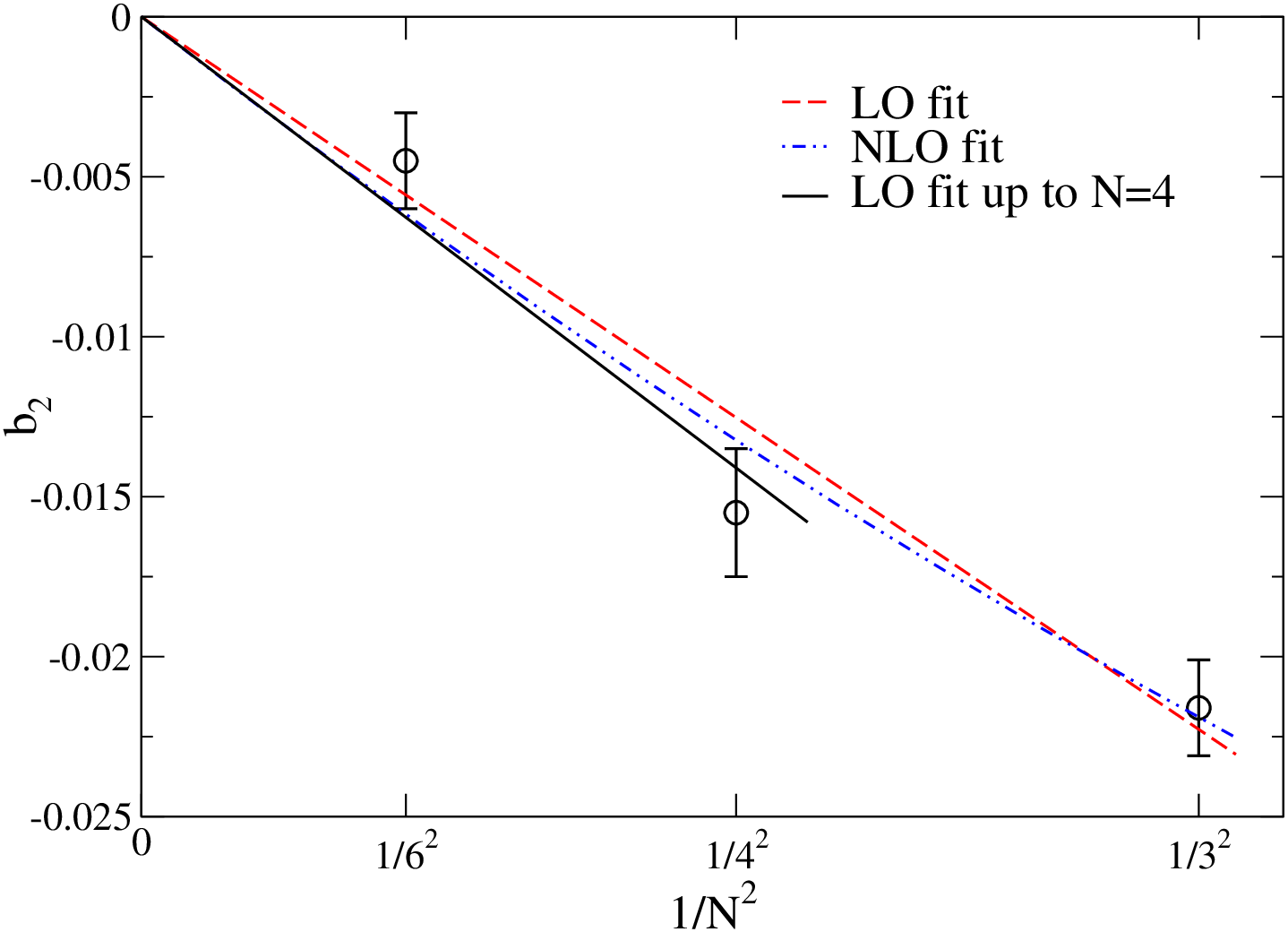}
\caption{Scaling of $b_2$ with the number of colors. Lines are result of a best
fit performed using the linear dependence expected from large $N$ arguments
(dashed line fitting all data, full line fitting only those for $N=4,\,6$), and
adding also a quadratic contribution (dotted-dashed line).}
\label{fig:b2largeN}
\end{figure}

The estimates of $b_2$ versus the number of colors are shown in
Fig.~\ref{fig:b2largeN}. They decrease with increasing $N$, strongly
supporting a vanishing large-$N$ limit.  Fitting the data to the
Ansatz $b_2(N)=c/N^\kappa$ we obtain $\kappa=1.9(3)$, fully supporting
the $1/N^2$ scaling predicted by the large-$N$ scaling arguments.

We now analyze the data assuming the $1/N^2$ scaling. Some fits are shown in
Fig.~\ref{fig:b2largeN}.  The leading form $b_2=\bar{b}_2/N^2$ of the expected
$N$ dependence is used with two different fit ranges: in one case all the data
are fitted, which gives $\bar{b}_2=-0.200(12)$ (with
$\chi^2/\mathrm{dof}\sim 2.9/2$), while in the other case only data with $N>3$
are used, obtaining $\bar{b}_2=-0.23(3)$ (with $\chi^2/\mathrm{dof}\sim
1.9/1$).  These results are in perfect agreement with those of the fit
performed using also the NLO correction, i.e. to $b_2=\bar{b}_2/N^2 +
\bar{b}_{2}^{(1)}/N^4$, that gives $\bar{b}_2=-0.23(5)$ and
$\bar{b}_2^{(1)}=0.3(5)$ (with $\chi^2/\mathrm{dof}\sim 2.5/1$),
further indicating the absence of significant NLO correction. As our final
estimate we report
\begin{equation}\label{barb2}
\bar{b}_2=-0.23(3)\ .
\end{equation} 
The previous estimate for this quantity in the literature was
$\bar{b}_2=-0.21(5)$ from \cite{DelDebbio:2002xa} and it should be
stressed that not only the error of the final result gets reduced in
the present study, but also the whole analysis is now much more solid,
since the old result relied heavily on the $SU(3)$ result.  

\begin{figure}
\includegraphics[width=0.92\columnwidth, clip]{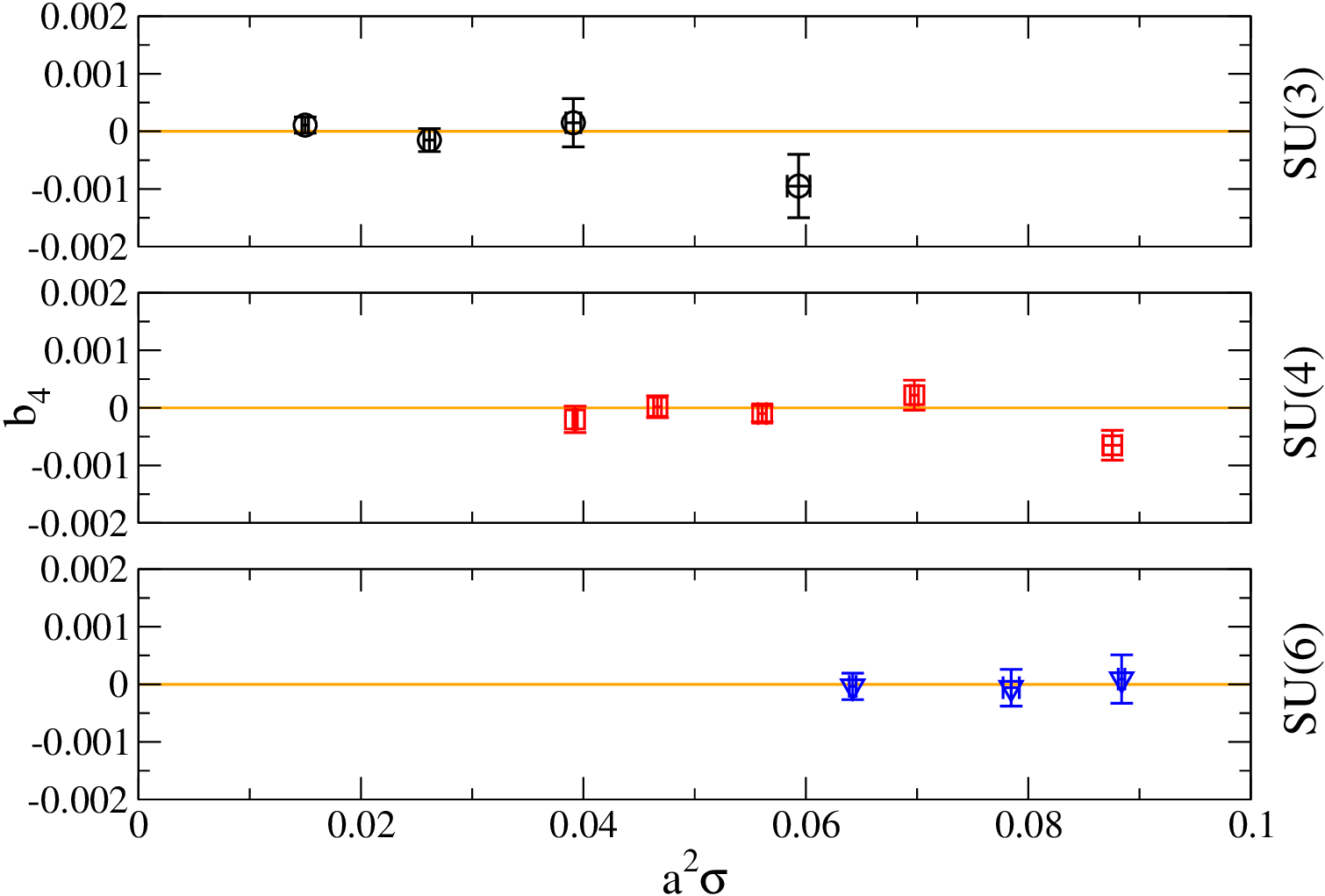}
\caption{Estimates of $b_4$ for $N=3,\,4,\,6$.}
\label{fig:b4allN}
\end{figure}

Some estimates of the $O(\theta^6)$ coefficient $b_4$ of the ground-state
energy density are shown in Fig.~\ref{fig:b4allN}.  To extract a
continuum value the same procedure adopted for $b_2$ was used also in this
case: linear fits in $a^2$ were performed and consistency with the results
obtained by discarding the values of the coarsest and the finest lattice
spacings was verified.  The final results are reported in
Table~\ref{tab:contvalues}.  As previously anticipated, they are still
compatible with zero.  Assuming the large-$N$ scaling $b_4\simeq \bar{b}_4/N^4$
for $N=4$, we obtain the bound
\begin{equation}\label{barb4}
|\bar{b}_4|\lesssim 0.1 \ .
\end{equation}

Our results for the large-$N$ coefficients $\bar{b}_{2n}$ may be compared with
the analytical calculations by holographic approaches~\cite{Witten:1998uka,
Parnachev:2008fy, Bigazzi:2014qsa, Bigazzi:2015bna}.  In particular, a
compatible (negative sign) result for $\bar{b}_2$ is reported in
Ref.~\cite{Bigazzi:2015bna}.

\begin{figure}
\includegraphics[width=0.92\columnwidth, clip]{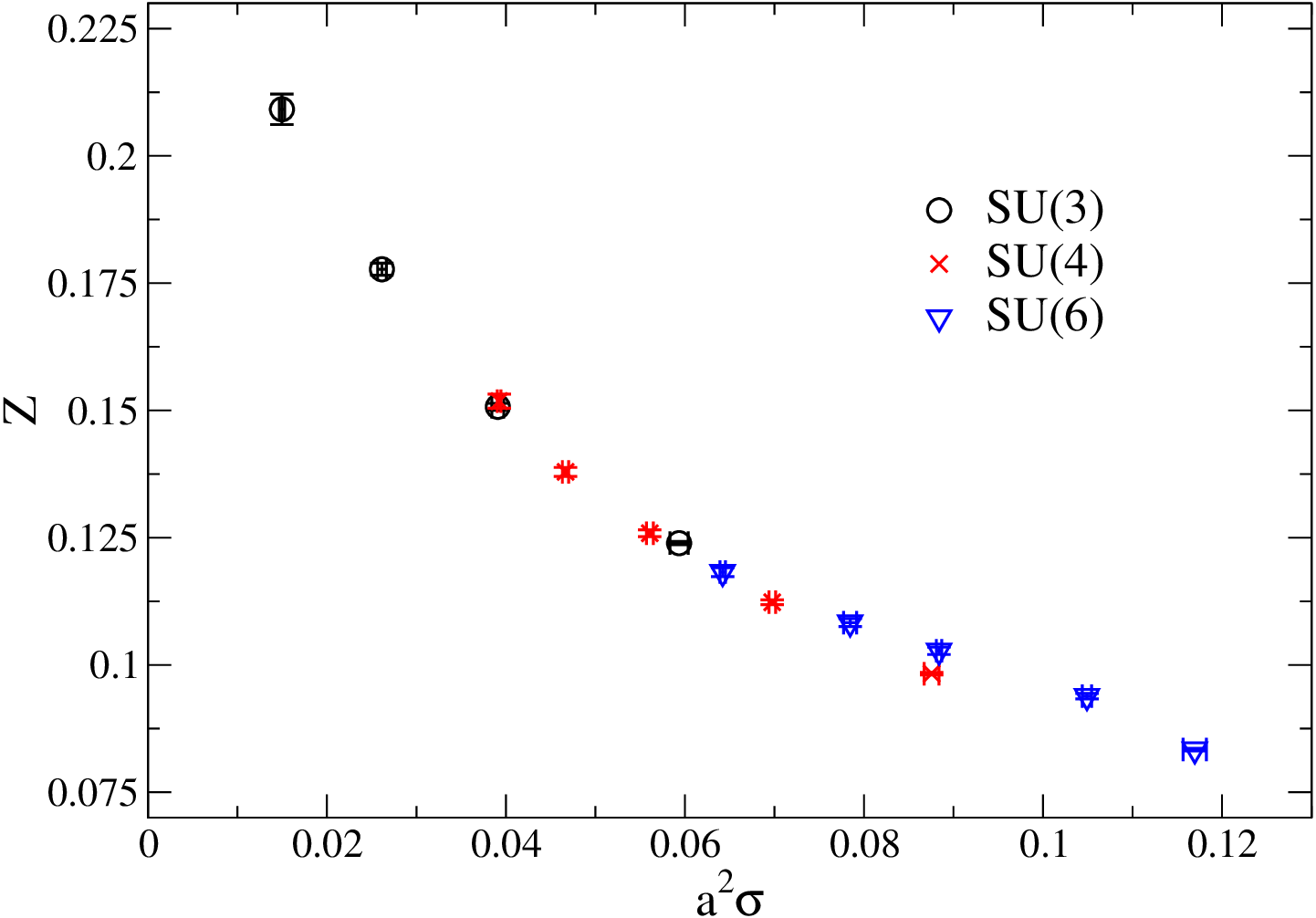}
\caption{Dependence of the renormalization constant $Z$ values on the
  lattice spacing for the case of three, four and six colors.}
\label{fig:ZallN}
\end{figure}

Finally in Fig.~\ref{fig:ZallN} we present our determinations of the
renormalization factor $Z$ for $N=3,4$ and $6$ and for the various
lattice spacings used (again $SU(3)$ data come from
\cite{Bonati:2015sqt}). It can be noted that all the data
approximately collapse on a common curve, i.e. $Z$ at fixed lattice
spacing has a well defined large-$N$ limit. This behaviour could have
been guessed by noting that the perturbative computation of $Z$
performed in \cite{Campostrini:1988cy} is in fact (up to subleading
corrections) an expansion in the 't Hooft coupling $g^2N$.

\section{Large $N$ in 2D $CP^{N-1}$ models}\label{sec:cpn}

The 2D $CP^{N-1}$ (Euclidean) Lagrangian in the presence of a
$\theta$ term is:
\begin{equation}\label{eq1}
\mathcal{L}_{\theta}(z,{\bar z}) = {N \over 2\,f} D_\mu {\bar z} D_\mu z + 
i {\theta \over 2\pi} \epsilon_{\mu \nu} \partial_\mu A_\nu,
\end{equation}
where $z$ is an $N$-component complex vector satisfying ${\bar z}\,z =
1$, $D_\mu \equiv \partial_\mu + i\,A_\mu$ and $A_\mu \equiv i\, {\bar
  z} \,\partial_\mu z$.  In order to analyze the large-$N$ behavior of
the models one must introduce the Lagrange multiplier fields
$\lambda_\mu$ and $\alpha$ and perform a Gaussian integration, thus
obtaining the effective action
\begin{equation} \label{eq2}
\begin{aligned}
S_{\rm eff}(\lambda_\mu, \alpha) = & N\, {\rm Tr}\,\ln [-D_\mu D_\mu
  +i\,\alpha] - \frac{N}{2\,f} \int \rmd^2 x \,[i\,\alpha] \\ & -i
\frac{\theta}{4\pi} \int \rmd^2x \,\, \epsilon_{\mu \nu} F_{\mu
  \nu}\ ,
\end{aligned}
\end{equation}
where now $D_\mu \equiv \partial_\mu +i\,\lambda_\mu$ and $F_{\mu \nu}
\equiv \partial_\mu \lambda_\nu - \partial_\nu \lambda_\mu$.  The
multiplier fields become dynamical and in particular $\lambda_\mu$
develops a massless pole, thus behaving as a bona fide (Abelian) gauge
field.

The functional evaluation of $E(\theta) - E(0)$ in the large-$N$ limit
can now be performed starting from the computation of the effective
potential $N\,V(A,B)$ as a function of the constant vacuum expectation
values $A(B)\equiv \langle i\,\alpha \rangle$ and $\langle F_{\mu \nu}
\rangle = \epsilon_{\mu \nu} B$. In Ref~\cite{Rossi:2016uce} it has
been shown that
\begin{equation}\label{eq3}
\begin{aligned}
V(A,B; \theta) = \frac{1}{4\,\pi} \Bigl[&- A\,\ln \frac{2 B}{m^2} 
-2 B\,\ln \Gamma \left(\frac{1}{2} +\frac{A}{2 B} \right) \\
&+B \ln 2\pi - 2\,i\,\frac{\theta}{N} \,B \Bigr] \ ,
\end{aligned}
\end{equation}
where $m^2\equiv A(B\to 0)$ and $\Gamma$ is the standard Gamma
function. It is now apparent that the natural expansion parameter for
the large-$N$ evaluation of $E$ is $\bar \theta \equiv \theta/N$
\cite{Witten:1980sp, Witten:1998uka}.

To the purpose of evaluating $E(\theta)$ one must then solve the
saddle point equations
\begin{equation}\label{eq4}
\frac{\partial V}{\partial A} = 0 \ , \qquad 
\frac{\partial V}{\partial B} = 0\ .
\end{equation}
The first equation may be employed in order to find the function
$A(B^2)$, independent of $\theta$, and to generate the large $N$
effective Lagrangian for the gauge degrees of freedom $V_{\lambda} (B;
\theta) \equiv V[A(B^2),B; \theta]$.

The dependence on $\bar \theta$ of the large $N$ vacuum energy can
now be found immediately from the relationship
\begin{equation}\label{eq5}
E(\bar \theta) - E(0) = N\,V_{\lambda} \bigl[B(\bar \theta),\bar\theta\bigr]\ ,
\end{equation}
where $B(\bar \theta)$ is the solution of the equation 
\begin{equation}
\frac{\partial V_{\lambda}}{\partial B} = 0\ .
\end{equation}
One must appreciate that solving the last equation implies a
continuation from real to complex values of $B$, that can be easily
performed in the perturbative regime by observing that $V_{\lambda}
(B; 0)$ admits an asymptotic expansion in the even powers of
$B$. Therefore it is possible to find a solution for purely imaginary
$B$ in the form of a power series in the odd powers of $\bar \theta$.

The first few terms of the expansion of $B(\bar \theta)$ are
\begin{equation}\label{btheta}
B(\bar \theta) \approx  6\, i\, m^2 \, \bar \theta \bigl(1- {54 \over 5}\bar \theta^2 
-{76014 \over 175} \bar \theta^4+...\bigr),
\end{equation}
where $m^2 = A(\theta=0)$ is a square mass scale.  Beside the leading
large-$N$ behavior of the topological
susceptibility~\cite{D'Adda:1978uc,Luscher:1978rn,Witten:1978bc,Campostrini:1991kv}
\begin{equation}
\chi \approx {3 m^2 \over \pi N},
\label{chiln}
\end{equation}
we obtain the rescaled coefficients $\bar b_{2n}\equiv 
{\rm lim}_{N\to\infty} N^{2n} b_{2n}$
of the $\theta$ expansion of the ground-state energy density:
\begin{eqnarray}
&& \bar{b}_2=-{27\over 5}, \\
&& \bar{b}_4=-\frac{25338}{175},                \nonumber  \\
&& \bar{b}_6=-{16198389\over 875}, \nonumber\\           
&& \bar{b}_8=-\frac{1500696182646}{336875},       \nonumber
\end{eqnarray}
etc...  These results for $b_{2n}$ extend those reported in
Ref.~\cite{DelDebbio:2006yuf} (in particular they correct the value of
$\bar{b}_4$).

An analysis of several higher order coefficients shows that they are
all negative and grow very rapidly, as one might have expected as a
consequence of the nonanalytic dependence of the effective Lagrangian
on $B$ already observed in Ref.~\cite{Rossi:2016uce}. In turn this
phenomenon can be related to the fact that the full-fledged dependence
on $\theta$ of the vacuum energy for any finite value of $N$ must
exhibit a $2 \pi$ periodicity which disappears in the large $N$ limit,
thus implying a noncommutativity of the expansions and a vanishing
radius of convergence in the variable $\bar{\theta} \equiv \theta/N$.

We finally mention that the large-$N$ behavior (\ref{chiln}) of the
topological susceptibility has been confirmed by numerical results of
lattice $CP^{N-1}$
models~\cite{CRV-92,V-92,ACDP-00,DMV-04,Vicari:2008jw}.  Instead,
numerical results for the $\theta$-expansion coefficients $b_{2n}$
have never been obtained yet.

\section{Conclusions}\label{sec:concl}

We study the large-$N$ scaling behavior of the $\theta$ dependence of
4D $SU(N)$ gauge theories and 2D $CP^{N-1}$ models, where $\theta$ is
the parameter associated with the Lagrangian topological term.  In
particular, we focus on the first few coefficients $b_{2n}$ of the
expansion (\ref{thdep}) of their ground-state energy $E(\theta)$
beyond the quadratic approximation, which parametrize the deviations
from a simple Gaussian distribution of the topological charge at
$\theta=0$.

We present a numerical analysis of Monte Carlo simulations of 4D
$SU(N)$ lattice gauge theories for $N=3,\,4,\,6$ in the presence of an
imaginary $\theta$ term. This method, based on the analytic
continuation of the $\theta$ dependence from imaginary to real
$\theta$ values, allows us to significantly improve earlier
determinations of the first few coefficients $b_{2n}$.  The results
provide a robust evidence of the large-$N$ behavior predicted by
standard large-$N$ scaling arguments, i.e., $b_{2n}= O(N^{-2n})$.  In
particular, we obtain $b_2=\bar{b}_2/N^2 + O(1/N^4)$ with
$\bar{b}_2=-0.23(3)$.  The results for the next coefficient $b_4$ of
the $\theta$ expansion (\ref{thdep}) show that it is very small, in
agreement with the large-$N$ prediction that $b_4=O(N^{-4})$. Assuming
the large-$N$ scaling $b_4\approx \bar b_4/N^4$, we obtain the bound
$|\bar b_4|\lesssim 0.1$.

An important issue concerns the consistency between the $\theta/N$
dependence in the large-$N$ limit and the $2\pi$ periodicity related
to the topological phase-like nature of $\theta$.  Indeed, the
large-$N$ scaling behavior is apparently incompatible with the
periodicity condition $E(\theta) = E(\theta+2\pi)$, which is a
consequence of the quantization of the topological charge, as
indicated by semiclassical arguments based on its geometrical meaning
for continuous field configurations~\cite{Gross:1980br}.  Indeed a regular function of
$\bar\theta=\theta/N$ cannot be invariant for $\theta \to
\theta+2\pi$, unless it is constant.  A plausible way
out~\cite{Witten:1980sp} is that the ground-state energy $E(\theta)$
tends to a multibranched function in the large-$N$ limit, such as
\begin{equation}
E(\theta) - E(0) = N^2 \, {\rm Min}_k\, H\left( {\theta+2\pi k\over
  N}\right),
\label{conj1}
\end{equation}
where $H$ is a generic function. $E(\theta)$ is then periodic in
$\theta$, but not regular everywhere.  As a consequence, the physical
relevance of the large-$N$ scaling of the $\theta$ dependence should
be only restricted to the power-law expansion (\ref{thdep}) around
$\theta=0$, and of analogous expansions of other observables, thus to
the $N$ dependence of their coefficients.

Our results significantly strengthen the evidence of the large-$N$
scaling scenario of the $\theta$ dependence, extending it beyond the
$O(\theta^2)$ expansion.  We note that the large-$N$ scaling of the
$\theta$ expansion is not guaranteed.  Indeed there are some notable
cases in which this does not apply. For example this occurs in the
high-temperature regime of 4D $SU(N)$ gauge theories: for high
temperatures the dilute instanton-gas approximation (DIGA) is expected to
provide reliable results and one gets (see e.g.  \cite{Gross:1980br})
the result $b_2=-1/12$ for any $N$ value. While the DIGA approximation
is a priori expected to be valid only at asymptotically high
temperatures, the switch from the large $N$ behavior to the instanton
gas behavior occurs at the deconfinement transition temperature $T_c$
\cite{Bonati:2013tt}.

\begin{figure}[t]
\includegraphics[width=0.92\columnwidth, clip]{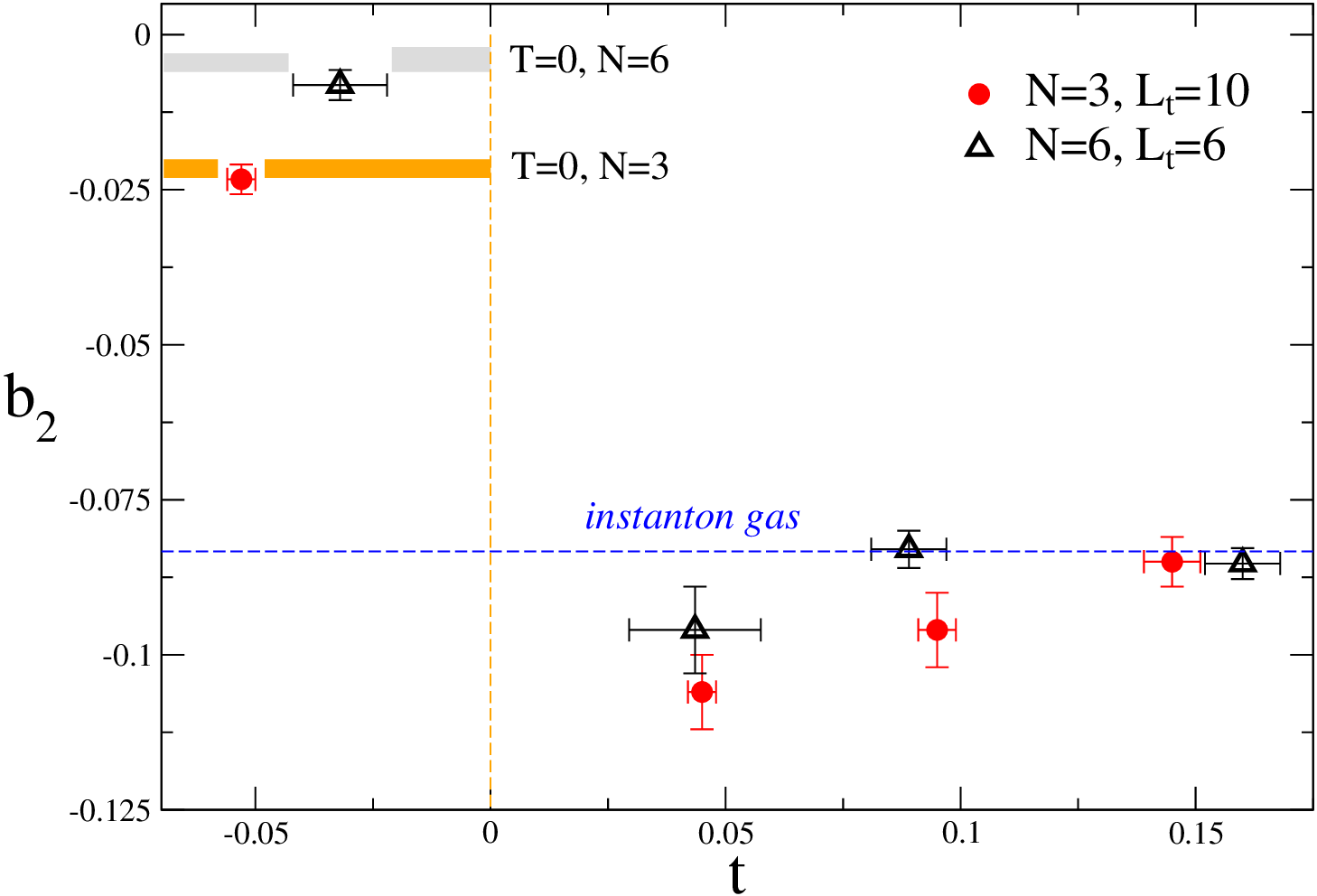}
\caption{Behaviour of $b_2$ across the deconfinement transition for
  $SU(3)$ and $SU(6)$ ($t$ is the reduced temperature defined by
  $t=(T-T_c)/T_c$).  The horizontal bands denote the zero temperature
  values. Updated version of the figure originally presented in
  \cite{Bonati:2013tt}.}
\label{fig:b2T}
\end{figure}

The analytic continuation method that we used to compute the $\theta$
dependence can be also exploited in finite-temperature simulation,
where it is typically even more efficient\footnote{Some caution is
  only needed for temperatures slightly above deconfinement, since the
  introduction of an imaginary $\theta$ term increases the critical
  temperature \cite{D'Elia:2012vv, D'Elia:2013eua}.}. As an example of
its application in finite-temperature runs, Fig.~\ref{fig:b2T}
presents an updating of the results presented in \cite{Bonati:2013tt}
regarding the change of $\theta$ dependence across the deconfinement
transition. While the results for $T>T_c$ were precise enough also in
the original publication, the region below deconfinement is much more
difficult (see the discussion in \cite{Bonati:2015sqt}). By combining
the result for $SU(3)$ obtained in \cite{Bonati:2015sqt} and the
present ones for $SU(6)$, in the left side of Fig.~\ref{fig:b2T} we
can now display the continuum extrapolated zero temperature value of
$b_2$ for $SU(6)$ and much more precise results for the finite
temperature values of $b_2$. These results confirm the results of
\cite{Bonati:2013tt} to an higher accuracy: in the low-temperature
phase the $\theta$-dependence properties, thus $\chi$ and $b_{2n}$,
appear almost temperature independent, up to an abrupt change across
the finite-temperature deconfinement transition.  Then, in the
high-temperature phase the $\theta$ dependence turns out to be that
predicted by DIGA, with $b_{2n}$ not depending on $N$.

Finally, this paper also reports a study of the large-$N$ $\theta$
dependence of the 2D $CP^{N-1}$ models, whose leading behavior can be
computed analytically. The results confirm the predicted large-$N$
scaling behavior $b_{2n}\approx \bar b_{2n} N^{-2n}$ for the
coefficients of the expansion of the ground-state energy around
$\theta=0$.

\acknowledgments We acknowledge useful discussions with Francesco
Bigazzi and Haris Panagopoulos.  Numerical simulations have been
performed on the Galileo machine at CINECA (under INFN project NPQCD),
on the CSN4 cluster of the Scientific Computing Center at INFN-PISA
and on GRID resources provided by INFN.

\appendix

\section{Cooling and gradient flow}\label{sec:cgf}
 
\begin{figure}
\includegraphics[width=0.92\columnwidth, clip]{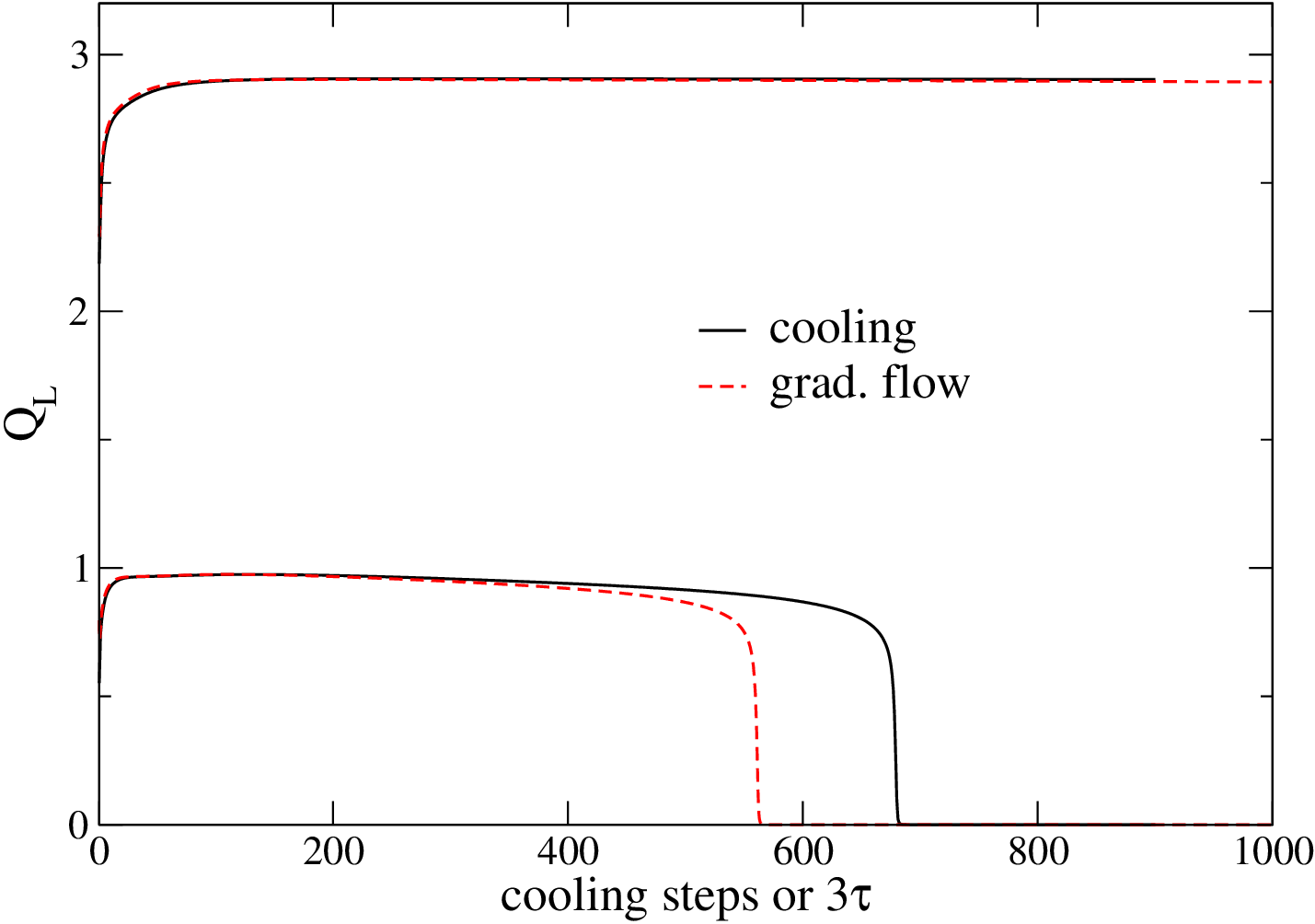}
\caption{Comparison of cooling and gradient flow evolutions for two $SU(6)$
configurations.}\label{cgf_fig}
\end{figure}

It was shown in \cite{Bonati:2014tqa} that cooling and the gradient
flow with Wilson action give identical results for the topological
charge when the number of cooling steps $n_c$ is related to the
dimensionless flow time $\tau$ by the relation $n_c=3\tau$. This
relation was explicitly verified by simulation in $SU(3)$ gauge theory
and it was later extended to improved gauge actions
\cite{Alexandrou:2015yba}. During the early stages of this work we
numerically verified on a subsample of configurations that, as
theoretically expected, the same relation holds true also in the
$SU(6)$ case. An example of the comparison between the two methods is
reported in Fig.~\ref{cgf_fig}, which displays some generic features:
in $SU(6)$ the topological charge is much more stable than in $SU(3)$,
to reach a plateau of $Q_L$ around 100 cooling steps are needed, for
very prolongated smoothing both cooling and gradient flow evolutions
tunnel to the topologically trivial configuration and the tunneling
typically happens first for the gradient flow.

\section{Parallel tempering in $\theta$}\label{sec:pt}

Parallel tempering \cite{Hukushima}, also know as replica exchange
Monte Carlo, is the most widely used variant of the simulated
tempering algorithm \cite{Marinari:1992qd} and was originally
introduced to speed up simulations of spin glasses. In this appendix
we report the results of some tests performed to investigate the
effectiveness of parallel tempering to reduce the autocorrelation of
the topological charge in $SU(6)$.

Parallel tempering is typically used in systems with complicated
energy landscapes to reduce the autocorrelation times.  The original
idea is to perform standard simulations at various temperatures (with
higher temperatures decorrelating faster than the lower ones) and once
in a while try to exchange the configurations at different
temperatures with a Metropolis-like step, that guarantees the detailed
balance and hence the stochastic exactness of the algorithms. In this
way the quickly decorrelating runs ``feed'' the slow ones and
autocorrelations are drastically reduced.

For the case of gauge theories the first natural choice would be to
use parallel tempering between runs at different $\beta$ values, with
the runs at large values of $\beta$ playing the role of the slowly
decorrelating ones.  Although from a theoretical point of view this
should work, one is faced with an efficiency problem: in order for the
exchanges to be accepted with reasonable probability the $\beta$
values have to be close to each other, in fact closer and closer as
the volume is increased, thus making the algorithm not convenient
apart from extreme cases. See e.g. Ref.~\cite{V-92} for applications
to the 2D $CP^{N-1}$ models.  This is the reason why alternative
procedures have been proposed to work with different $\beta$ values,
that are closer in spirit to the idea of multi-level simulations, see
e.g.  \cite{Endres:2015yca}.

\begin{figure}
\includegraphics[width=0.92\columnwidth, clip]{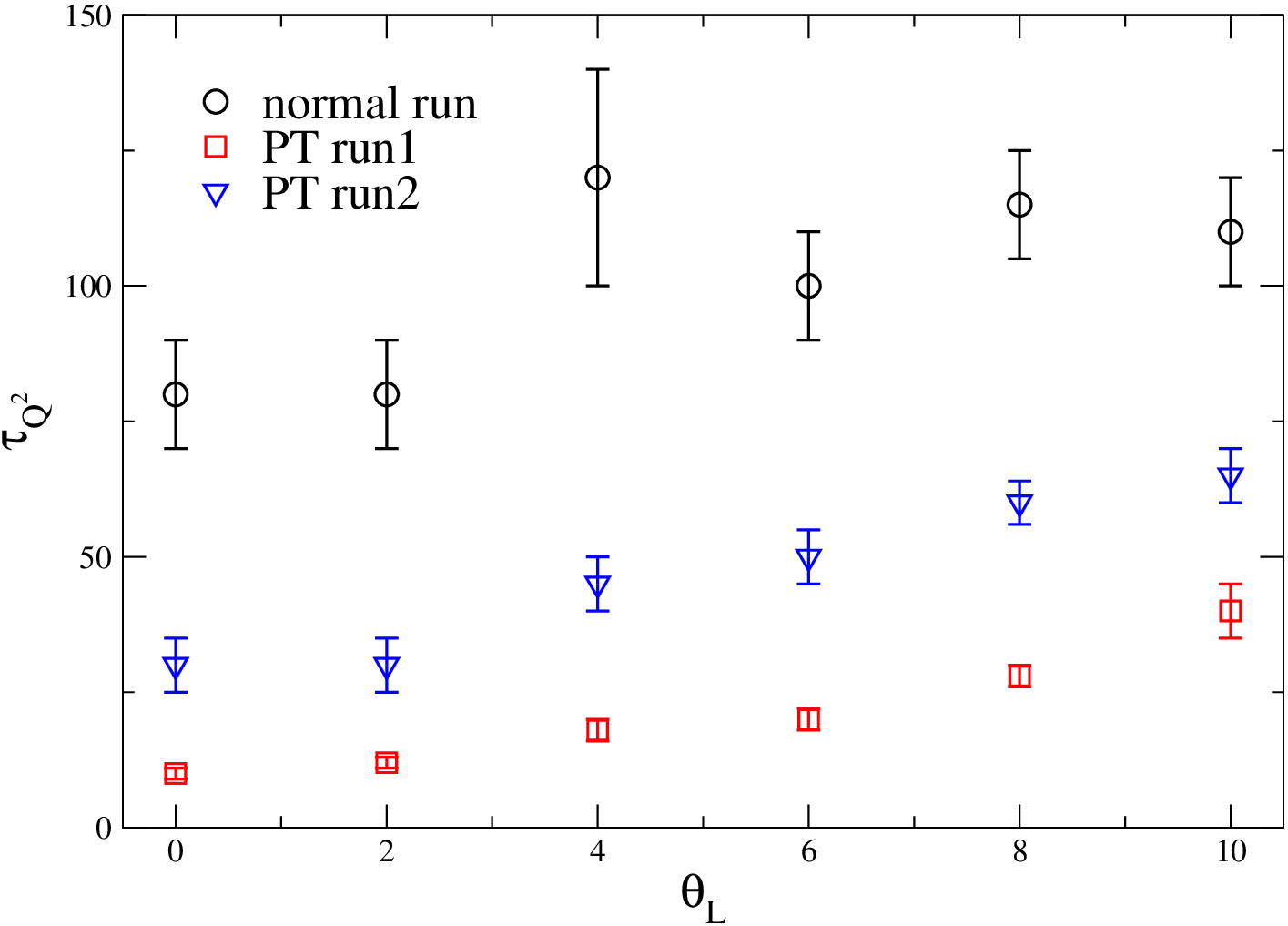}
\caption{Autocorrelation times (in units of measure) of the square of the
topological charge for the standard run and for the two tests with parallel
tempering. In $run1$ an exchange was proposed every 4 measures, while in $run2$
it was proposed every 40 measures.}
\label{fig:pt}
\end{figure}

Since we are using simulations at nonvanishing values of the $\theta$
angle, an alternative possibility is to perform the switch step of the
parallel tempering between runs at different $\theta_L$ values
\cite{Panagopoulos:2011rb,thetapt}. In this case there are no ``fast''
and ``slow'' runs, but since the mean values of the topological charge
are different for different $\theta_L$ values, the switch step
characteristic of the parallel tempering is expected to effectively
increase the tunneling rate of the topological charge.

As a testbed for the parallel tempering in $\theta_L$ we used $SU(6)$
with coupling $\beta=25.056$ and $\theta_L$ values from $-10$ to $10$
with $\Delta\theta_L=2$. Using the standard algorithm described in
Sec.~\ref{sec:numset} the autocorrelation time of the square of the
topological charge is around $100$ measures (with 1 measure every 10
updates) and we tried two different exchange frequencies in the
parallel tempering: in the run denoted by $run1$ an exchange was
proposed every 4 measures, while in $run2$ it was proposed every 40
measures; in both the cases the proposed switch was accepted with a
probability of about $70\%$.

\begin{table*}[t]
\begin{tabular}{l|l|l|l|l|l|l|l|l}
\hline $\beta$ & $L/a$ & $a\sqrt{\sigma}$ & $\tau_{Q^2}$ & $Z$         & $a^4\chi$                & $\chi/\sigma^2$ & $b_2$ & $b_4$ \\
\hline \rule{0mm}{3mm}10.720  & 12  & 0.2959(14)       & 0.80(5)      & 0.09828(26) & $2.296(11) \times 10^{-4}$ & 0.02995(58) & -0.01628(62) & -0.00065(26) \\
\hline \rule{0mm}{3mm}10.816  & 12  & 0.2642(7)        & 1.6(1)       & 0.11231(49) & $1.4152(72)\times 10^{-4}$ & 0.02905(34) & -0.01658(79) &  \phantom{ }0.00022(26) \\
\hline \rule{0mm}{3mm}10.912  & 12  & 0.2368(6)        & 2.5(5)       & 0.12586(66) & $8.971(67) \times 10^{-5}$ & 0.02853(36) & -0.01617(67) & -0.00010(14) \\
\hline \rule{0mm}{3mm}11.008  & 14  & 0.2160(8)        & 6.0(5)       & 0.13792(88) & $6.044(48) \times 10^{-5}$ & 0.02776(47) & -0.01526(84) &  \phantom{ }0.00002(19) \\
\hline \rule{0mm}{3mm}11.104  & 16  & 0.1981(5)        & 14(1)        & 0.1518(14)  & $4.129(62) \times 10^{-5}$ & 0.02681(48) & -0.0167(11)  & -0.00020(23) \\
\hline
\end{tabular}
\caption{$SU(4)$ data. String tension data from
  \cite{DelDebbio:2001sj}.  Autocorrelation times of the square of the
  topological charge are expressed in unit of measurements (one
  measure every 10 updates, see Sec.~\ref{sec:numset} for more details
  on the update) and have been evaluated using the blocking method.}
\label{tabsu4}
\end{table*}

\begin{table*}[t]
\begin{tabular}{l|l|l|l|l|l|l|l|l}
\hline $\beta$ & $L/a$ & $a\sqrt{\sigma}$ & $\tau_{Q^2}$ & $Z$ & $a^4\chi$ & $\chi/\sigma^2$ & $b_2$ & $b_4$ \\

\hline \rule{0mm}{3mm} 24.500 &  12 & 0.3420(19)*      & 1.8(2)       &  0.08338(25) & $3.870(12)\times 10^{-4}$ & 0.02828(63) &  -0.01030(92) &  \phantom{ }0.00008(60)  \\ 
\hline \rule{0mm}{3mm} 24.624 &  10 & 0.3239(8)        & 4.2(3)       &  0.09386(54) & $2.654(13)\times 10^{-4}$ & 0.02412(27) &  -0.0096(10)  &  \phantom{ }0.00011(31)  \\
\hline \rule{0mm}{3mm} 24.768 &  12 & 0.2973(5)        & 11(1)        &  0.10278(73) & $1.856(13)\times 10^{-4}$ & 0.02375(23) &  -0.0056(13)  &  \phantom{ }0.00009(42)  \\ 
\hline \rule{0mm}{3mm} 24.845 &  12 & 0.2801(13)*      & 22(3)        &  0.10832(78) & $1.545(11)\times 10^{-4}$ & 0.02509(50) &  -0.0049(11)  &  -0.00006(32)            \\
\hline \rule{0mm}{3mm} 25.056 &  12 & 0.2534(6)        & 80(10)       &  0.11822(85) & $9.770(68)\times 10^{-5}$ & 0.02370(28) &  -0.0047(10)  &  -0.00004(23)            \\
\hline 
\end{tabular}
\caption{$SU(6)$ data. Most of the string tension data came from
  \cite{DelDebbio:2001sj}, these denoted by * from
  \cite{Lucini:2003zr}.  Autocorrelation times of the square of the
  topological charge are expressed in unit of measurements (one
  measure every 10 updates, see Sec.~\ref{sec:numset} for more details
  on the update) and have been evaluated using the blocking method.}
\label{tabsu6}
\end{table*}

The autocorrelation times of $Q^2$ for the different values of
$\theta_L$ and the various run are shown in Fig.~\ref{fig:pt}. As was
to be expected given the range of $\theta_L$ used in the parallel
tempering, small $\theta_L$ runs decorrelate faster than the ones with
large $\theta_L$, and in all the cases an important decrease of
$\tau_{Q^2}$ is observed, that is more significant for the case of
$run1$, in which exchanges were proposed at higher rate than in
$run2$.  In the best case the autocorrelation time was reduced by
around an order of magnitude with respect to the standard runs.

With respect to the single run at $\theta_L=0$ this reduction of
$\tau_{Q^2}$ is however not sufficient to compensate for the CPU time
required to perform the update of the 11 replicas used in the parallel
tempering, since simulations at nonvanishing $\theta_L$ values are
about 2.5 more time consuming than simulation at $\theta_L=0$.

On the other hand, the idea of the method of analytic continuation in
$\theta$ for computing the $b_{2n}$ coefficients is exactly to use
several $\theta_L$ values anyway, so that one can still hope to have
an efficiency gain. This is however not the case: the simulations
performed at different $\theta_L$ values are obviously correlated in
the parallel tempering and, taking this correlation into account, no
gain is apparently obtained by using the parallel tempering in the
computation e.g. of $b_2$.

A possible explanation of this result (i.e. strong reduction of the
autocorrelation for the single $\theta_L$ run and strong correlation
between different $\theta_L$ runs) is the following. While on average
the lattice operator $Q_L$ is obviously related to the operator $Q$,
the specific form of their UV fluctuations can be different and are
larger, in particular, for $Q_L$.  As a consequence, the Metropolis
test for the exchange of configurations, which is solely based on
$Q_L$, could be easier, but then not accompained by a fast
decorrelation of the global topological content $Q$ after the
exchange, which would proceed with a decorrelation time likely
comparable with the $\tau_{Q^2}$ of the standard simulation. If this
interpretation is correct, then the observed reduction of the
autocorrelation times at fixed $\theta_L$ is just a consequence of the
reshuffling of the configurations induced by the exchanges, which are
very frequent due to the largest UV fluctuations of $Q_L$. The update
of the global information contained in the time histories at different
$\theta_L$ values, which is the one used in the global fit, suffers
instead from the usual autocorrelation problems.

One possibility, in order to improve the performance of the parallel
tempering algorithm, could be to adopt an improved discretization of
$Q_L$, e.g. a smeared definition of the topological 
charge density, such as those considered in 
Refs.~\cite{CDPV-96,Laio:2015era}; this would require to abandon the
heatbath and overrelaxation algorithms in favour of an Hybrid Monte
Carlo approach~\cite{Duane:1987de}.  However, it is not clear a priori
whether that would result in an improvement of the global
decorrelation properties, i.e. in a final net gain, or rather in a
deterioration of the autocorrelation time for the single trajectory at
fixed $\theta_L$, because of the rarer configuration reshuffling.

\section{Numerical data}\label{sec:data}

In Tab.~\ref{tabsu4} and Tab.~\ref{tabsu6} we report the data obtained for
$SU(4)$ and $SU(6)$ respectively at the different values of the coupling
studied, together with the values of the string tension used in the analysis.

\end{document}